\documentstyle [10pt,amsfonts] {article}
\input epsf

\newcommand{\beq}{\begin{equation}}
\newcommand{\eeq}{\end{equation}}
\newcommand{\beqs}{\begin{eqnarray}}
\newcommand{\eeqs}{\end{eqnarray}}
\newcommand{\lsim}{\mathrel{\raisebox{-.6ex}{$\stackrel{\textstyle<}{\sim}$}}}
\newcommand{\gsim}{\mathrel{\raisebox{-.6ex}{$\stackrel{\textstyle>}{\sim}$}}}

\catcode`@=11
\@addtoreset{equation}{section}
\@addtoreset{equation}{subsection}
\def\theequation{\ifnum\value{section}=0 \arabic{equation}\ignorespaces
\else \ifnum\value{section}=-1 A.\arabic{equation}\ignorespaces
\else \ifnum\value{subsection}=0 \thesection.\arabic{equation}\ignorespaces
\else \thesection.\arabic{subsection}.\arabic{equation}\ignorespaces
                           \fi
                      \fi
                 \fi}
\catcode`@=12

\begin{document}

\def\thefootnote{\fnsymbol{footnote}}

\baselineskip 5.0mm

\vspace{4mm}

\begin{center}

{\Large \bf Exact Potts Model Partition Functions on Wider Arbitrary-Length 
Strips of the Square Lattice}

\vspace{8mm}

\setcounter{footnote}{0}
Shu-Chiuan Chang\footnote{email: shu-chiuan.chang@sunysb.edu} and 
\setcounter{footnote}{6}
Robert Shrock\footnote{email: robert.shrock@sunysb.edu}

\vspace{6mm}

C. N. Yang Institute for Theoretical Physics  \\
State University of New York       \\
Stony Brook, N. Y. 11794-3840  \\

\vspace{10mm}

{\bf Abstract}
\end{center}

We present exact calculations of the partition function of the $q$-state Potts
model for general $q$ and temperature on strips of the square lattice of width
$L_y=3$ vertices and arbitrary length $L_x$ with periodic longitudinal boundary
conditions, of the following types: (i) $(FBC_y,PBC_x)=$ cyclic, (ii)
$(FBC_y,TPBC_x)=$ M\"obius, (iii) $(PBC_y,PBC_x)=$ toroidal, and (iv)
$(PBC_y,TPBC_x)=$ Klein bottle, where $FBC$ and $(T)PBC$ refer to free and
(twisted) periodic boundary conditions.  Results for the $L_y=2$ torus and
Klein bottle strips are also included. In the infinite-length limit the 
thermodynamic properties are discussed and some general results are
given for low-temperature behavior on strips of arbitrarily great width.  We
determine the submanifold in the ${\mathbb C}^2$ space of $q$ and temperature
where the free energy is singular for these strips.  Our calculations are also
used to compute certain quantities of graph-theoretic interest.

\vspace{16mm}

\pagestyle{empty}
\newpage

\pagestyle{plain}
\pagenumbering{arabic}
\renewcommand{\thefootnote}{\arabic{footnote}}
\setcounter{footnote}{0}

\section{Introduction}

The $q$-state Potts model has served as a valuable model for the study of phase
transitions and critical phenomena \cite{potts,wurev}.  On a lattice, or, more
generally, on a (connected) graph $G$, at temperature $T$, this model is 
defined by the partition function
\beq
Z(G,q,v) = \sum_{ \{ \sigma_n \} } e^{-\beta {\cal H}}
\label{zfun}
\eeq
with the (zero-field) Hamiltonian
\beq
{\cal H} = -J \sum_{\langle i j \rangle} \delta_{\sigma_i \sigma_j}
\label{ham}
\eeq
where $\sigma_i=1,...,q$ are the spin variables on each vertex (site) 
$i \in G$;
$\beta = (k_BT)^{-1}$; and $\langle i j \rangle$ denotes pairs of adjacent
vertices.  The graph $G=G(V,E)$ is defined by its vertex set $V$ and its edge
set $E$; we denote the number of vertices of $G$ as $n=n(G)=|V|$ and the
number of edges of $G$ as $e(G)=|E|$.  We use the notation
\beq
K = \beta J \ , \quad a = u^{-1} = e^K \ , \quad v = a-1
\label{kdef}
\eeq
so that the physical ranges are (i) $a \ge 1$, i.e., $v \ge 0$ corresponding to
$\infty \ge T \ge 0$ for the Potts ferromagnet, and (ii) $0 \le a \le 1$,
i.e., $-1 \le v \le 0$, corresponding to $0 \le T \le \infty$ for the Potts
antiferromagnet.  One defines the (reduced) free energy per site $f=-\beta F$,
where $F$ is the actual free energy, via
\beq
f(\{G\},q,v) = \lim_{n \to \infty} \ln [ Z(G,q,v)^{1/n}] 
\label{ef}
\eeq
where we use the symbol $\{G\}$ to denote $\lim_{n \to \infty}G$ for a given
family of graphs.  In the present context, this $n \to \infty$ limit
corresponds to the limit of infinite length for a strip graph of the square 
lattice of fixed width and some prescribed boundary conditions. 

Let $G^\prime=(V,E^\prime)$ be a spanning subgraph of $G$, i.e. a subgraph
having the same vertex set $V$ and an edge set $E^\prime \subseteq E$. Then
$Z(G,q,v)$ can be written as the sum \cite{birk}-\cite{kf}
\beqs
Z(G,q,v) & = & \sum_{G^\prime \subseteq G} q^{k(G^\prime)}v^{e(G^\prime)}
\label{cluster} \cr\cr\cr
& = & \sum_{r=k(G)}^{n(G)}\sum_{s=0}^{e(G)}z_{rs} q^r v^s
\label{zpol}
\eeqs 
where $k(G^\prime)$ denotes the number of connected components of
$G^\prime$ and $z_{rs} \ge 0$.  Since we only consider connected graphs $G$, we
have $k(G)=1$. The formula (\ref{cluster}) enables one to generalize $q$ from
${\mathbb Z}_+$ to ${\mathbb R}_+$ (keeping $v$ in its physical range).  The
formula (\ref{cluster}) shows that $Z(G,q,v)$ is a polynomial in $q$ and $v$
(equivalently, $a$) with maximum and minimum degrees indicated in
eq. (\ref{zpol}).  The Potts model partition function on a graph $G$ is
essentially equivalent to the Tutte polynomial \cite{tutte1}-\cite{tutte3} and
Whitney rank polynomial \cite{whit}, \cite{wurev}, \cite{bbook}-\cite{boll} for
this graph, as discussed in the appendix.

In this paper we shall present exact calculations of the $q$-state Potts model
partition function $Z(G,q,v)$ for arbitrary $q$ and temperature for strips of
the square lattice of width $L_y=3$ vertices and arbitrary length $L_x$, with
periodic longitudinal boundary conditions of the following types: (i)
$(FBC_y,PBC_x)=$ cyclic, (ii) $(FBC_y,TPBC_x)=$ M\"obius, (iii)
$(PBC_y,PBC_x)=$ toroidal, and (iv) $(PBC_y,TPBC_x)=$ Klein bottle, where $FBC$
and $(T)PBC$ refer to free and (twisted) periodic boundary conditions,
respectively, and the longitudinal and transverse directions are taken to be
$\hat x$ (horizontal) and $\hat y$ (vertical).  Results for the $L_y=2$ torus
and Klein bottle strips are also included for comparison.  This is an extension
of our previous exact calculations of Potts model partition functions for
$L_y=2$ strips of the square lattice \cite{bcc,a} and other lattices
\cite{ta}-\cite{hca} including cyclic and M\"obius boundary
conditions.  (Results for free longitudinal boundary conditions are in 
\cite{a}-\cite{hca} and \cite{ks}, \cite{ts}.) 

There are several motivations for these exact calculations of Potts model
partition functions for strips of the square lattice. Clearly, new exact
calculations of Potts model partition functions are of value in their own
right.  From these, in the limit of infinite-length strips, one derives exact
thermodynamic functions and can make rigorous comparisons of various properties
among strips of different lattices.  In particular, with these exact results
one can study both the $T=0$ critical point of the $q$-state Potts ferromagnet
and $T=0$ properties of the Potts antiferromagnet. Further, via the
correspondence with the Tutte polynomial, our calculations yield several
quantities of relevance to mathematical graph theory.

Various special cases of the Potts model partition function are of interest.
One special case is the zero-temperature limit of the Potts antiferromagnet. 
For sufficiently large $q$, on a given lattice or graph $G$, this 
exhibits nonzero ground state entropy (without frustration).  This is of
interest as an exception to the third law of thermodynamics \cite{al,cw}. 
This is equivalent to a ground
state degeneracy per site (vertex), $W > 1$, since $S_0 = k_B \ln W$.  The
$T=0$ (i.e., $v=-1$) partition function of the above-mentioned $q$-state Potts
antiferromagnet on $G$ satisfies
\beq 
Z(G,q,-1)=P(G,q)
\label{zp}
\eeq
where $P(G,q)$ is the chromatic polynomial (in $q$) expressing the number
of ways of coloring the vertices of the graph $G$ with $q$ colors such that no
two adjacent vertices have the same color \cite{birk,bbook,rrev,rtrev}. The 
minimum number of colors necessary for this coloring is the chromatic number
of $G$, denoted $\chi(G)$.  We have 
\beq 
W(\{G\},q)= \lim_{n \to \infty}P(G,q)^{1/n} \ .
\label{w}
\eeq
A subtlety in the definition due to noncommutativity 
limits
\beq
\lim_{q \to q_s} \lim_{n \to \infty} P(G,q)^{1/n} \ne \lim_{n \to
\infty} \lim_{q \to q_s}P(G,q)^{1/n}
\label{wnoncom}
\eeq
at certain points $q_s$ was discussed in \cite{w}.

As derived in \cite{a}, a general form for the Potts model partition function
for the strip graphs $G_s$ considered here, or more generally, for recursively
defined families of graphs comprised of $m$ repeated subunits (e.g. the columns
of squares of height $L_y$ vertices that are repeated $L_x=m$ times to form an
$L_x \times L_y$ strip of a regular lattice with some specified boundary
conditions), is 
\beq 
Z((G_s)_m,q,v) = \sum_{j=1}^{N_{Z,G_s,\lambda}} c_{Z,G_s,j} 
(\lambda_{Z,G_s,j}(q,v))^m
\label{zgsum}
\eeq
where $N_{Z,G_s,\lambda}$, $c_{Z,G_s,j}$, and $\lambda_{Z,G_s,j}$ depend on 
the type of recursive family $G_s$ 
(lattice structure and boundary conditions) but not on its length $m$. 
For special case of the $T=0$ antiferromagnet, the partition
function, or equivalently, the chromatic polynomial $P((G_s)_m,q)$ has the
corresponding form 
\beq
P((G_s)_m,q) = \sum_{j=1}^{N_{P,G_s,\lambda}} c_{P,G_s,j} 
(\lambda_{P,G_s,j}(q))^m \ .
\label{pgsum}
\eeq
For sufficiently large integer $q$ values, the coefficients in the chromatic
polynomial (\ref{pgsum}) can be interpreted as multiplicities of eigenvalues 
of coloring matrices $T_P$ \cite{matmeth}, and the
sums of these coefficients are thus sums of dimensions of invariant subspaces
of these matrices.  An analogous statement applies to the coefficients in the
Potts model partition function (\ref{zgsum}), in terms of a generalized
coloring matrix $T_Z$. For a given family $G_s$ of strip graphs we shall 
denote these sums as 
\beq
C_{Z,G_s}=\sum_{j=1}^{N_{Z,G_s,\lambda}} c_{Z,G_s,j}
\label{czsum}
\eeq
and
\beq
C_{P,G_s}=\sum_{j=1}^{N_{P,G_s,\lambda}} c_{P,G_s,j} \ .
\label{cpsum}
\eeq
Below we shall give results for the determinants of the coloring matrices;
following the notation in \cite{cf} we have
\beq
\det T_Z(G_s) = \prod_{j=1}^{N_{Z,G_s,\lambda}} 
(\lambda_{Z,G_s,j})^{c_{Z,G_s,j}}
\label{detformz}
\eeq
and
\beq
\det T_P(G_s) = \prod_{j=1}^{N_{P,G_s,\lambda}} 
(\lambda_{P,G_s,j})^{c_{P,G_s,j}} \ .
\label{detformp}
\eeq
Let us define the shorthand notation
\beq
D_P(G_s)=\det T_P(G_s) \ , \quad D_Z(G_s)=\det T_Z(G_s) \ .
\label{dnotation}
\eeq

Here we distinguish the quantities $N_{Z,G_s,\lambda}$, $N_{P,G_s,\lambda}$; 
$c_{Z,G_s,j}$, $c_{P,G_s,j}$; $\lambda_{Z,G_s,j}$, and $\lambda_{P,G_s,j}$.  
Below, for
brevity of notation, we shall sometimes suppress the $Z$ or $P$ subscripts 
and the $s$ subscript on $G_s$ when
the meaning is clear from context and no confusion will result.

Using the formula (\ref{cluster}) for $Z(G_s,q,v)$, one can generalize $q$ from
${\mathbb Z}_+$ not just to ${\mathbb R}_+$ but to ${\mathbb C}$ and $a$ from
its physical ferromagnetic and antiferromagnetic ranges $1 \le a \le \infty$
and $0 \le a \le 1$ to $a \in {\mathbb C}$.  A subset of the zeros of $Z$ in
the two-complex dimensional space ${\mathbb C}^2$ defined by the pair of
variables $(q,a)$ can form an accumulation set in the $n \to \infty$ limit,
denoted ${\cal B}$, which is the continuous locus of points where the free
energy is nonanalytic.  This locus occurs where there is a non-analytic
switching between different $\lambda_{Z,G_s,j}$ of maximal magnitude and is
thus determined as the solution to the equation of degeneracy in magnitude of
these maximal or dominant $\lambda_{Z,G_s,j}$'s \cite{bcc,a}.  For a given
value of $a$, one can consider this locus in the $q$ plane, and we denote it as
${\cal B}_q(\{G\},a)$.  In the special case $a=0$ (i.e., $v=-1$) where the
partition function is equal to the chromatic polynomial, the zeros in $q$ are
the chromatic zeros, and ${\cal B}_q(\{G\},a=0)$ is their continuous
accumulation set in the $n \to \infty$ limit.  In previous papers we have given
exact calculations of the chromatic polynomials and nonanalytic loci ${\cal
B}_q$ for various families of graphs.  Some early related papers are 
\cite{bds}-\cite{klein}. Our methods for calculating these
chromatic polynomials were discussed in \cite{w},\cite{strip}-\cite{hs} and
some relevant bounds were given in \cite{b},\cite{ww}-\cite{sb}.  
The families of of 
graphs of primary interest here are strips of regular lattices with periodic 
longitudinal boundary conditions; previous related papers on such families of
graphs are \cite{bds}, \cite{w}, \cite{pg}-\cite{tor4}.  
With the exact Potts partition function for arbitrary temperature, one can
study ${\cal B}_q$ for $a \ne 0$ and, for a given value of $q$, one can study
the continuous accumulation set of the zeros of $Z(G,q,v)$ in the $a$ plane;
this will be denoted ${\cal B}_a(\{G\},q)$.  It will often be convenient to
consider the equivalent locus in the $u=1/a$ plane, namely ${\cal
B}_u(\{G\},q)$.  We shall sometimes write ${\cal B}_q(\{G\},a)$ simply as
${\cal B}_q$ or ${\cal B}$ when $\{G\}$ and $a$ are clear from the context, and
similarly with ${\cal B}_{a}$ and ${\cal B}_{u}$.  One gains a unified
understanding of the separate loci ${\cal B}_q(\{G\})$ for fixed $a$ and ${\cal
B}_a(\{G\})$ for fixed $q$ by relating these as different slices of the locus
${\cal B}$ in the ${\mathbb C}^2$ space defined by $(q,a)$ as we shall do here.
As discussed earlier \cite{bcc,a}, in the infinite-length limit where the locus
${\cal B}$ is defined, for a given width and transverse boundary condition
(free or periodic) ${\cal B}$ depends on the longitudinal boundary conditions
but is independent of whether they involve orientation reversal or not.  Thus,
in the present context, for a given $L_y$, ${\cal B}$ is the same for the 
cyclic and M\"obius strips, and separately for the torus and Klein bottle
strips. 

Following the notation in \cite{w} and our other earlier works on ${\cal
B}_q(\{G\})$ for $a=0$, we denote the maximal region in the complex $q$ plane
to which one can analytically continue the function $W(\{G\},q)$ from physical
values where there is nonzero ground state entropy as $R_1$ .  The maximal
value of $q$ where ${\cal B}_q$ intersects the (positive) real axis was
labelled $q_c(\{G\})$.  Thus, region $R_1$ includes the positive real axis for
$q > q_c(\{G\})$.  Correspondingly, in our works on complex-temperature
properties of spin models \cite{chisq} we had labelled the 
complex-temperature extension
(CTE) of the physical paramagnetic phase as (CTE)PM, which will simply be
denoted PM here, the extension being understood, and similarly with
ferromagnetic (FM) and antiferromagnetic (AFM); other complex-temperature
phases, having no overlap with any physical phase, were denoted $O_j$ (for
``other''), with $j$ indexing the particular phase \cite{chisq,chihc}.  Here we
shall continue to use this notation for the respective slices of ${\cal B}$ in
the $q$ and $a$ or $u$ planes. Various special cases of $Z(G,q,v)$ applicable
for arbitrary graphs $G$ were given in \cite{a}.  Just as was true for the
zero-temperature antiferromagnet, for the full partition function, at 
certain special points $q_s$ (typically $q_s=0,1...,\chi(G)$) one has
\beq 
\lim_{n \to \infty} \lim_{q
\to q_s} Z(G,q,v)^{1/n} \ne \lim_{q \to q_s} \lim_{n \to \infty} Z(G,q,v)^{1/n}
\ .
\label{fnoncomm}
\eeq
Because of
this noncommutativity, the formal definition (\ref{ef}) is, in general,
insufficient to define the free energy $f$ at these special points $q_s$; it is
necessary to specify the order of the limits that one uses in eq.
(\ref{fnoncomm}) \cite{w,a}. 

The Potts ferromagnet has a zero-temperature phase transition in the $L_x \to
\infty$ limit of the strip graphs considered here, and this has the consequence
that for cyclic and M\"obius boundary conditions, ${\cal B}$ passes through the
$T=0$ point $u=0$.  It follows that ${\cal B}$ is noncompact in the $a$ plane.
Hence, it is usually more convenient to study the slice of ${\cal B}$ in the
$u=1/a$ plane rather than the $a$ plane.  For the ferromagnet, since $a \to
\infty$ as $T \to 0$ and $Z$ diverges like $a^{e(G_s)}$ in this limit, we shall
use the reduced partition function $Z_r$ defined by
\beq
Z_r((G_s)_m,q,v)=a^{-e((G_s)_m)}Z((G_s)_m,q,v)=u^{e((G_s)_m)}Z((G_s)_m,q,v)
\label{zr}
\eeq
which has the finite limit $Z_r \to q$ as $T \to 0$.  For a general strip
graph $(G_s)_m$ of type $G_s$ and length $L_x=m$, we can write
\beqs
Z_r((G_s)_m,q,v) & = & u^{e((G_s)_m)}\sum_{j=1}^{N_{Z,G_s,\lambda}} c_{Z,G_s,j}
(\lambda_{Z,G_s,j})^m \equiv \sum_{j=1}^{N_{Z,G_s,\lambda}} c_{Z,G_s,j}
(\lambda_{Z,G_s,j,u})^m
\label{zu}
\eeqs
with
\beq
\lambda_{Z,G_s,j,u}=u^{e((G_s)_m)/m}\lambda_{Z,G_s,j} \ .
\label{lamu}
\eeq
For the Potts model partition functions of the width $L_y$ strips of the square
lattice with (i) cyclic or M\"obius and (ii) torus or Klein bottle boundary
conditions, the prefactor in eq. (\ref{lamu}) is (i) $u^{2L_y-1}$, and (ii) 
$u^{2L_y}$, respectively.

For the following, it will be convenient to define some general functions.
First, we define the following polynomial:
\beq
D_k(q) = \frac{P(C_k,q)}{q(q-1)} = 
\sum_{s=0}^{k-2}(-1)^s {{k-1}\choose {s}} q^{k-2-s}
\label{dk}
\eeq
and $P(C_n,q)$ is the chromatic polynomial for the circuit
(cyclic) graph $C_n$ with $n$ vertices, given by
\beq
P(C_n,q) = (q-1)^n +(q-1)(-1)^n \ . 
\label{pcn}
\eeq

Second, we define polynomials of degree $d$ in $q$ which are related to
Chebyshev polynomials:
\beq
c^{(d)}=U_{2d}\Bigl (\frac{\sqrt{q}}{2} \Bigr )
\label{cd}
\eeq
where $U_n(x)$ is the Chebyshev polynomial of the second kind, defined
by 
\beq
U_n(x) = \sum_{j=0}^{[\frac{n}{2}]} (-1)^j {n-j \choose j} (2x)^{n-2j}
\label{undef}
\eeq
where in eq. (\ref{undef}) the notation $[\frac{n}{2}]$ in the upper limit on 
the summand means the integral part of $\frac{n}{2}$.  These polynomials enter
as coefficients in (\ref{zgsum}) and (\ref{pgsum}) for the cyclic and M\"obius
strips under study here. The first few of these coefficients are 
\beq
c^{(0)}=1 \ , \quad c^{(1)}=q-1 \ ,
\label{cd01}
\eeq
\beq
c^{(2)}=q^2-3q+1,
\label{cd2}
\eeq
and
\beq
c^{(3)}=q^3-5q^2+6q-1 \ .
\label{cd3}
\eeq

The ``falling factorial'' used in combinatorics is defined as
\beq
q_{(n)} = \prod_{s=0}^{n-1}(q-s)
\label{ff}
\eeq
(with $q_{(0)} \equiv 1$), so that $q_{(1)}=q$, $q_{(2)}=q(q-1)$,
$q_{(3)}=q(q-1)(q-2)$, etc. 

On each transverse slice of the width $L_y$ cyclic and M\"obius strips for $L_y
\ge 3$, the $(L_y-2)$ interior vertices have degree (=coordination number) 4,
while the two exterior vertices have degree 3.  Let us define an effective 
degree or coordination number $\Delta_{eff}$ as 
\beq
\Delta_{eff} = \lim_{n \to \infty}\Bigl ( \frac{2|E|}{n} \Bigr ) \ .
\label{deltaeff}
\eeq
Then
\beq
\Delta_{eff}=4\Bigl ( 1 - \frac{1}{2L_y} \Bigr ) \quad {\rm for} \ \ 
sq, L_y, FBC_y,(T)PBC_x  \ . 
\label{delta_eff_cyc}
\eeq 
Thus, for $L_y=3$, $\Delta_{eff}=10/3$.  The strips of the square
lattice with torus and Klein bottle boundary conditions are $\Delta$-regular 
graph with $\Delta=4$.  (Here, a $\Delta$-regular graph is defined as one in
which each vertex has the same degree, $\Delta$.) 
We proceed to describe our results.

\section{Potts Model Partition Function for Cyclic and M\"obius $L_y=3$ 
Strips of the Square Lattice}

\subsection{General Structural Properties}

For cyclic and M\"obius strips of the square lattice of fixed width $L_y$ and
arbitrary length $L_x$, the coefficients $c_{Z,G_s,j}$ and $c_{P,G_s,j}$ in
eqs. (\ref{zgsum}) and (\ref{pgsum}) depend only on $L_y$ and the boundary
conditions.  These coefficients are polynomials in $q$.  In the case of cyclic
boundary conditions, the coefficients of degree $d$ in $q$ have a unique form,
namely $c^{(d)}$ given in (\ref{cd}), while in the case of M\"obius boundary
conditions, the coefficients of degree $d$ are $\pm c^{(d)}$.  For the cyclic
strip, let us denote the number of terms $\lambda_{Z,G_s,j}$ and 
$\lambda_{P,G_s,j}$ in eqs. (\ref{zgsum}) and (\ref{pgsum}) respectively 
that have for their coefficient $c^{(d)}$, $n_Z(L_y,d)$ and $n_P(L_y,d)$. 
For the M\"obius strip, we denote the number of terms $\lambda_{Z,G_s,j}$ and
$\lambda_{P,G_s,j}$ in eqs. (\ref{zgsum}) and (\ref{pgsum}) 
that have for their coefficient $\pm c^{(d)}$, $n_Z(L_y,d,\pm)$ and 
$n_P(L_y,d,\pm)$.  In \cite{cf} we proved a number of general theorems on the
structure of $Z(G_s,q,v)$ and $P(G_s,q)$ that determined these numbers, as well
as the overall number of terms $N_{Z,G_s,\lambda}$ and $N_{P,G_s,\lambda}$. 
For our present purposes, we can apply the general formula \cite{cf}, 
\beq
N_{Z,L_y,\lambda}={2L_y \choose L_y}
\label{nztot}
\eeq
with $L_y=3$ to conclude that
\beq
N_{Z,3,\lambda}=20 \ .
\label{nztotly3}
\eeq
Further, \cite{cf} 
\beq
n_Z(L_y,d)=\frac{(2d+1)}{(L_y+d+1)}{2L_y \choose L_y-d}
\label{nzlyd}
\eeq
for $0 \le d \le L_y$, with $n_Z(L_y,d)=0$ for $d > L_y$; hence, for 
$L_y=3$, it follows that 
\beq
n_Z(3,0)=5 \ , n_Z(3,1)=9 \ , n_Z(3,2)=5 \ , n_Z(3,3)=1 \ .
\label{nzly3d}
\eeq
For the M\"obius strip, the analysis in \cite{cf} gives, for the special case
$L_y=3$ 
\beq
n_Z(3,0,+)=6 \ , n_Z(3,0,-)=4 \ , n_Z(3,1,+)=6 \ , n_Z(3,1,-)=3 \ , 
n_Z(3,2,+)=1
\label{nzlydmb}
\eeq
with other $n_Z(3,d,\pm)=0$. 

In the special case of the zero-temperature Potts antiferromagnet, $v=-1$, from
the general result \cite{cf}
\beq
N_{P,L_y,\lambda}=2(L_y-1)! \ \sum_{j=0}^{[\frac{L_y}{2}]} \frac{(L_y-j)}{
(j!)^2(L_y-2j)!}
\label{nptot}
\eeq
one has, for $L_y=3$, 
\beq
N_{P,3,\lambda}=10 \ .
\label{nptotly3}
\eeq
Further, for the $L_y=3$ cyclic and M\"obius strips, \cite{cf}
\beq
n_P(3,0)=2 \ , n_P(3,1)=4 \ , n_P(3,2)=3 \ , n_P(3,3)=1
\label{nply3}
\eeq
\beq
n_P(3,0,+)=3 \ , n_P(3,0,-)=2 \ , n_P(3,1,+)=3 \ , n_P(3,1,-)=1 \ , 
n_P(3,2,+)=1
\label{nplydmb}
\eeq
with other $n_P(3,d,\pm)=0$.

For the cyclic strip, (\ref{czsum}) and (\ref{cpsum}) are \cite{cf} 
\beq
C_{Z,L_y} = \sum_{d=0}^{L_y} n_Z(L_y,d)c^{(d)}=q^{L_y}
\label{czsumcyc}
\eeq
and
\beq
C_{P,L_y} = \sum_{d=0}^{L_y} n_P(L_y,d)c^{(d)}=P(T_{L_y},q)=q(q-1)^{L_y-1}
\label{cpdsumcyc}
\eeq
where $T_n$ denotes the path or tree graph with $n$ vertices.  Hence, 
for the $L_y=3$ cyclic strip under study here, 
\beq
C_{Z,3}=q^3 \ , \quad C_{P,3}=q(q-1)^2 \ .
\label{czcply3}
\eeq
The corresponding sums of coefficients for the M\"obius strips are, for the 
case of odd $L_y$ relevant here, 
\beq
C_{Z,L_y,Mb}=q^{(L_y+1)/2}
\label{czsummob}
\eeq
and
\beq
C_{P,L_y,Mb}=P(T_{(L_y+1)/2},q)=q(q-1)^{\frac{L_y-1}{2}}
\label{cpsummob}
\eeq
so that for $L_y=3$, 
\beq
C_{Z,3,Mb}=q^2 \ , \quad C_{P,3,Mb}=q(q-1) \ .
\label{czpsummob}
\eeq

\subsection{Results for the Potts Model Partition Function} 

Using a systematic iterative application of the deletion-contraction theorem
and transfer matrix methods, we have calculated the exact Potts model 
partition function $Z(G,q,v)$ for general $q$ and $v$ for the the cyclic 
and M\"obius strips of the square lattice of width $L_y=3$ and arbitrary 
length $L_x=m$.  For quantities that are independent of $L_x$, we shall use 
the abbreviations $s3c$ and $s3Mb$ for these strips.  We find for these 
partition functions the results 
\beq
Z(sq,3 \times m,FBC_y,PBC_x,q,v) = 
\sum_{j=1}^{20} c_{Z,s3c,j}(\lambda_{Z,s3c,j}(q,v))^m 
\label{zgsums3cyc}
\eeq
and
\beq
Z(sq,3 \times m,FBC_y,TPBC_x,q,v) = 
\sum_{j=1}^{20} c_{Z,s3Mb,j}(\lambda_{Z,s3c,j}(q,v))^m
\label{zgsums3mb}
\eeq
where (ordering the $\lambda_{Z,s3c,j}$'s by decreasing degree of the 
associated coefficients $c_{Z,s3c,j}$) 
\beq
\lambda_{Z,s3c,1}=v^3
\label{lam1}
\eeq
\beq
\lambda_{Z,s3c,2}=v^2(q+v)
\label{lam2}
\eeq
\beq
\lambda_{Z,s3c,(3,4)}=\frac{v^2}{2}\Bigl [ q+v(v+5) \pm \Bigl (
[q+v(v+5)]^2-4v(v+q)(v+1) \Bigr )^{1/2} \biggr ] 
\label{lam34}
\eeq
and
\beq
\lambda_{Z,s3c,(5,6)}=\frac{v^2}{2}\Bigl [q+v(v+3) \pm  \Bigl (
[q+v(v+3)]^2-4v(v+q)(v+1) \Bigr )^{1/2} \biggr ] \ . 
\label{lam56}
\eeq
Note that
\beq
\lambda_{Z,s3c,3}\lambda_{Z,s3c,4}=
\lambda_{Z,s3c,5}\lambda_{Z,s3c,6}=v^5(v+1)(v+q) \ . 
\label{lamcyc334}
\eeq
The terms $\lambda_{Z,s3c,j}$ for $j=7,8,9$ are roots of the cubic equation
\beq
\xi^3+f_{31}\xi^2+f_{32}\xi+f_{33}=0
\label{eqcub}
\eeq
where
\beq
f_{31}=-v(2v+q)(3v+q+v^2)
\label{f31}
\eeq
\beq
f_{32}=v^3(v+q)(v+1)(6v^2+4qv+q^2+v^3)
\label{f32}
\eeq
and
\beq
f_{33}=-v^6(v+q)^3(v+1)^2 \ . 
\label{f33}
\eeq
Further, 
\beq
\lambda_{Z,s3,16}=(v+q)(v+1)v^2 \ . 
\label{lam16}
\eeq
The terms $\lambda_{Z,s3c,j}$ for $10 \le j \le 15$ and $17 \le j \le 20$ are
roots of algebraic equations of degree 6 and 4, respectively, given as
(\ref{eqsix}) and (\ref{eqquart}) in the appendix.  The roots of the quartic 
equation (\ref{eqquart}) comprise the $\lambda_{Z,G,j}$ for the $3 \times L_x$
strip of the square lattice with $(FBC_y,FBC_x)$.  

For the cyclic strip the coefficients in eq. (\ref{zgsums3cyc}) are, in terms
of the $c^{(d)}$ given above in eq. (\ref{cd}), 
\beq
c_{Z,s3c,1}=c^{(3)}
\label{c1cyc}
\eeq
\beq
c_{Z,s3c,j}=c^{(2)} \quad {\rm for} \ \ 2 \le j \le 6
\label{c2to6cyc}
\eeq
\beq
c_{Z,s3c,j}=c^{(1)} \quad {\rm for} \ \ 7 \le j \le 15
\label{c7to15cyc}
\eeq
and
\beq
c_{Z,s3c,j}=c^{(0)} \quad {\rm for} \ \ 16 \le j \le 20 \ . 
\label{c16to20}
\eeq
For the M\"obius strip, the coefficients in (\ref{zgsums3mb}) are 
\beq
c_{Z,s3Mb,1}=c^{(2)}
\label{cd2_mb}
\eeq
\beq
c_{Z,s3Mb,j}=c^{(0)} \quad {\rm for} \quad j=5,6,17,18,19,20
\label{cd0_mb}
\eeq
\beq
c_{Z,s3Mb,j}=-c^{(0)} \quad  {\rm for} \ \ j=2,3,4,16
\label{cd0minus_mb}
\eeq
\beq
c_{Z,s3Mb,j}=c^{(1)} \quad {\rm for} \quad 10 \le j \le 15
\label{cd1_mb}
\eeq
\beq
c_{Z,s3Mb,j}=-c^{(1)} \quad {\rm for} \quad 7 \le j \le 9 \ .
\label{cd1minus_mb}
\eeq
One easily checks that these results agree with the general structural formulas
given above. 

 From our calculations of $Z(G,q,v)$ for both the $L_y=2$ and $L_y=3$ cyclic
and M\"obius strips of the square lattice, as well as the elementary
calculation of the cyclic $L_y=1$ case, we infer that the first two terms
are consistent with the following general formula for arbitrary $L_y$ and 
$G=sq,L_y \times L_x, FBC_y,(T)PBC_x$:
\beq
\lambda_{Z,G,1}=v^{L_y} 
\label{lam1general}
\eeq
\beq
\lambda_{Z,G,2} = v^{L_y-1}(q+v)
\label{lam2general}
\eeq
with coefficients
\beq
c_{Z,G_{cyc.},1}=c^{(L_y)}
\label{c1general}
\eeq
and
\beq
c_{Z,G_{cyc.},2}=c^{(L_y-1)}
\label{c2general}
\eeq
for the cyclic strips; for the M\"obius strips, the corresponding coefficients
can be determined by our general formulas in \cite{cf}.

 From these results we calculate the determinant
\beqs
D_Z(sq,3 \times L_x,FBC_y,PBC_x) & = & 
\prod_{j=1}^{20} (\lambda_{Z,s3c,j})^{c_{Z,s3c,j}} \cr\cr 
& = & v^{3q^2(q-1)}(v+1)^{2q^2}(v+q)^{3q^2} \cr\cr
& = & (y-1)^{3q^3}y^{2q^2}x^{3q^2}
\label{lamprodcyc3}
\eeqs
where the variables $x$ and $y$ are defined in (\ref{xdef}) and (\ref{ydef}) in
the appendix and are the variables of the Tutte polynomial. 
Comparing this with $D_Z(G)$ for the $L_y=1$ and $L_y=2$ strips \cite{a,cf}, 
we note that these cases $L_y=1,2,3$, are all described by the 
following formula
\beq
D_Z(sq,L_y \times L_x,FBC_y,PBC_x) = (y-1)^{L_y q^{L_y}}
y^{(L_y-1)q^{L_y-1}}x^{L_y q^{L_y-1}} \ . 
\label{lamprodcycly}
\eeq
We also find, for the M\"obius strip, 
\beqs
D_Z(sq,3 \times L_x,FBC_y,TPBC_x) & = & 
\prod_{j=1}^{20} (\lambda_{Z,s3c,j})^{c_{Z,s3Mb,j}} \cr\cr 
& = & v^{3q(q-1)}(v+1)^{2q}(v+q)^{3q}
\cr\cr
& = & (y-1)^{3q^2}y^{2q}x^{3q} \ . 
\label{lamprodmb3}
\eeqs
Comparing this with $D_Z(sq,2 \times L_x,FBC_y,TPBC_x)$ \cite{a,cf}, we 
observe that both of these cases $L_y=2,3$ are described by the formula
\beq
D_Z(sq,L_y \times L_x,FBC_y,PBC_x)=(y-1)^{L_y q^{L_y-1}}y^{(L_y-1)q}
x^{L_y q^{L_y-2}} \ . 
\label{lamprodmbly}
\eeq

\subsection{$v=-1$ Special Case} 

For the special case $v=-1$, i.e., the zero-temperature Potts antiferromagnet,
where the partition function reduces to the chromatic polynomial, ten of the 
twenty $\lambda_{Z,s3c,j}$'s vanish and the others reduce to those for the 
$L_y=3$ cyclic/M\"obius strip given in \cite{wcy,pm}.  To see which
$\lambda_{Z,s3c,j}$'s vanish and which reduce to corresponding
$\lambda_{P,s3c,j}$'s, we recall the expressions for the chromatic polynomials:
\beq
P(sq,3 \times m,FBC_y,PBC_x,q)\sum_{j=1}^{10} c_{P,s3c,j}(\lambda_{P,s3c,j})^m
\label{psumcyc}
\eeq
\beq
P(sq,3 \times m,FBC_y,TPBC_x,q)=\sum_{j=1}^{10} 
c_{P,s3Mb,j}(\lambda_{P,s3c,j})^m \ . 
\label{psummb}
\eeq
We order the terms $\lambda_{P,s3c,j}$ by decreasing degree of the associated
coefficients $c_{P,s3c,j}$ for the cyclic strip, as polynomials in $q$ (this
differs from the order and hence labelling used in \cite{wcy,pm}). These terms 
are
\beq
\lambda_{P,s3c,1}=-1
\label{lamp1}
\eeq
\beq
\lambda_{P,s3c,2}=q-1
\label{lamp2}
\eeq
\beq
\lambda_{P,s3c,3}=q-4
\label{lamp3}
\eeq
\beq
\lambda_{P,s3c,4}=q-2
\label{lamp4}
\eeq
and
\beq
\lambda_{P,s3c,5}=-(q-2)^2 \ . 
\label{lamp5}
\eeq
The terms $\lambda_{P,s3c,j}$, $j=6,7,8$, are the roots of the cubic equation
\beq
\xi^3+b_{31}\xi^2+b_{32}\xi+b_{33}=0
\label{sqcubic}
\eeq
with
\beq
b_{31}=2q^2-9q+12
\label{bsq21}
\eeq
\beq
b_{32}=q^4-10q^3+36q^2-56q+31
\eeq
\label{bsq22}
\beq
b_{33}=-(q-1)(q^4-9q^3+29q^2-40q+22) \ .
\label{bsq23}
\eeq
Finally, we have 
\beq
\lambda_{P,s3c,(9,10)} = \frac{1}{2}\Biggl [ (q-2)(q^2-3q+5) \pm
\Bigl \{ (q^2-5q+7)(q^4-5q^3+11q^2-12q+8) \Bigr \}^{1/2} \Biggr ] \ .
\label{lamp910}
\eeq
Thus, the correspondences are as follows.  For $v \to -1$, 
\beq
\lambda_{Z,s3c,1} \rightarrow \lambda_{P,s3c,1}
\label{lam1a0}
\eeq
and
\beq
\lambda_{Z,s3c,2} \rightarrow \lambda_{P,s3c,2} \ . 
\label{lam2a0}
\eeq
For the two pairs of quadratic roots, we have 
\beqs
\lambda_{Z,s3c,3} & \rightarrow & \lambda_{P,s3c,3} \cr\cr
\lambda_{Z,s3c,4} & \rightarrow & 0
\label{lam34a0}
\eeqs
and 
\beqs
\lambda_{Z,s3c,5} & \rightarrow & \lambda_{P,s3c,4} \cr\cr
\lambda_{Z,s3c,6} & \rightarrow & 0 \ .
\label{lam56a0}
\eeqs
For $v=-1$ the cubic equation (\ref{eqcub}) reduces to $\xi^2[\xi+(q-2)^2]=0$, 
so, with the indicated labelling of roots, 
\beqs
\lambda_{Z,s3c,7} & \rightarrow & \lambda_{P,s3c,5} \cr\cr
\lambda_{Z,s3c,j} & \rightarrow & 0 \ \ {\rm for} \ \ j=8,9 \ . 
\label{lam789a0}
\eeqs
The sixth order equation (\ref{eqsix}) reduces to $\xi^3$ times the cubic
equation (\ref{sqcubic}) so that, with the indicated labelling, 
\beqs
\lambda_{Z,s3c,j} & \rightarrow & \lambda_{P,s3c,j-4} \quad {\rm for} \ \ 
10 \le j \le 12 \cr\cr
\lambda_{Z,s3c,j} & \rightarrow & 0  \quad {\rm for} \ \ 13 \le j \le 16 \ . 
\label{lam1016a0}
\eeqs
Finally, the quartic equation (\ref{eqquart}) becomes $\xi^2$ times a
quadratic, yielding the reduction 
\beqs
\lambda_{Z,s3c,j} & \rightarrow & \lambda_{P,s3c,j-8} \quad {\rm for} \ \ 
j=17,18 \cr\cr
\lambda_{Z,s3c,j} & \rightarrow & 0 \quad {\rm for} \ \
j=19,20 \ . 
\label{lam1720}
\eeqs
The sums and products of the eigenvalues of the coloring matrix for the $L_y=3$
chromatic polynomials of the 
cyclic and M\"obius strips were given in \cite{cf}. 
Note that, since some of the eigenvalues of the coloring matrix for the full
Potts model vanish when $v=-1$, one must extract these from the expressions
(\ref{lamprodcyc3}) and (\ref{lamprodmb3}) in order to recover the
products for the corresponding chromatic polynomials.  For example, for the
cyclic $L_y=3$ case, one has \cite{cf} 
\beqs
& & \prod_{j=1}^{10} (\lambda_{P,s3c,j})^{c_{P,s3c,j}}
= (-1)^{(q-2)(q^2-3q+1)}(q-1)^{(q-1)^2}
(q-2)^{q^2-q-1} \times \cr\cr
& & (q-4)^{q^2-3q+1}(q^3-6q^2+13q-11)(q^4-9q^3+29q^2-40q+22)^{q-1}
\label{dsqly3p}
\eeqs

\subsection{$v=0$ Special Case} 

For infinite temperature, $v=0$, we have 
\beq
\lambda_{Z,s3c,17}=q^3  \quad {\rm for} \ \ v=0
\label{lam5v0}
\eeq
\beq
\lambda_{Z,s3c,j}=0 \quad {\rm for} \ \ v=0 \ \ {\rm and} \ \ j \ne 17
\label{lamne5v0}
\eeq
so that the partition function reduces to 
$Z(sq,3 \times m,FBC_y,(T)PBC_x,q,v=0)=q^n$.  This can be seen directly from
eqs. (\ref{zfun}) and (\ref{ham}) or from eq. (\ref{zpol}).

\section{Potts Model Partition Function for $L_y=2$ and $L_y=3$ Strips of the 
Square Lattice with Torus and Klein Bottle Boundary Conditions}

\subsection{General Structural Properties}

For the strip with torus boundary conditions, (\ref{czsum}) and (\ref{cpsum}) 
are \cite{cf} 
\beq
C_{Z,L_y} = q^{L_y}
\label{czsumtorus}
\eeq
as in (\ref{czsumcyc}), and
\beq
C_{P,L_y} = P(C_{L_y},q)
\label{cpdsumtor}
\eeq
where $C_n$ is the circuit graph and $P(C_n,q)$ was given above in (\ref{pcn}).
Thus, for the $L_y=2$ and $L_y=3$ strips with toroidal boundary conditions 
under study here, 
\beq
C_{Z,2}=q^2 \ , \quad C_{P,2}=q_{(2)}=q(q-1)
\label{czcply2tor}
\eeq
and
\beq
C_{Z,3}=q^3 \ , \quad C_{P,3}=q_{(3)}=q(q-1)(q-2) \ . 
\label{czcply3tor}
\eeq
The corresponding sums of coefficients for the width $L_y$ strips with Klein 
bottle boundary conditions (denoted $k$) are
\beq
C_{Z,L_y,k}=q^{[\frac{L_y+1}{2}]}
\label{czsumklein}
\eeq
where $[\nu]$ denotes the integral part of $\nu$, and \cite{cf}
\beq
C_{P,L_y,k}=0
\label{cpsumklein}
\eeq
so that for $L_y=2,3$, 
\beq
C_{Z,2,k}=q \ , \quad C_{Z,3,k}=q^2, 
\label{czsumklein23}
\eeq
and 
\beq
C_{P,2,k}=C_{P,3,k}=0 \ . 
\label{cpsumklein23}
\eeq

\subsection{Results for the Potts Model Partition Function}

By the same methods as for the cyclic and M\"obius strips, we have calculated
the exact Potts model partition function $Z(G_s,q,v)$ for general $q$ and $v$
for the the strips of the square lattice of width $L_y=3$, arbitrary length
$L_x=m$, and torus or Klein bottle boundary conditions.  For comparison, we
also include our calculation of the partition function for the corresponding
width $L_y=2$ strips. For quantities that are independent of $L_x$, we shall
use the abbreviations $s2t$, $s2k$, $s3t$, and $s3k$ for these strips.

\subsection{$L_y=2$}

Strictly speaking, the $L_y=2$ strips of the square lattice with torus or Klein
bottle boundary conditions are not proper graphs but multigraphs, i.e., they
involve multiple (double) edges connecting vertical pairs of vertices.  The
number of vertices and edges are given by $n=L_xL_y$ and $e=4L_x$.  The graphs
in the family are $\Delta$-regular with $\Delta=4$.
We find for the partition function the results 
\beq
N_{Z,s2t,\lambda}=N_{Z,s2k,\lambda}=6
\label{nztottorus2}
\eeq
\beq
Z(sq,2 \times m,PBC_y,PBC_x,q,v) =
\sum_{j=1}^6 c_{Z,s2t,j}(\lambda_{Z,s2t,j}(q,v))^m
\label{zgsums2torus}
\eeq
and 
\beq
Z(sq,2 \times m,PBC_y,TPBC_x,q,v) =
\sum_{j=1}^6 c_{Z,s2k,j}(\lambda_{Z,s2t,j}(q,v))^m
\label{zgsums2klein}
\eeq
where
\beq
\lambda_{Z,s2t,1}=v^2
\label{lamtor2_1}
\eeq
\beq
\lambda_{Z,s2t,2}=v(v+q)
\label{lamtor2_2}
\eeq
\beq
\lambda_{Z,s2t,(3,4)} = \frac{v}{2}\Bigl [ q+6v+4v^2+v^3 \pm \sqrt{R_{234}} \ 
\Bigr ]
\label{lamtor2_34}
\eeq
where
\beq
R_{234}=32v^2+q^2+8qv+40v^3+8v^5+v^6-2qv^3+24v^4
\label{r234}
\eeq
\beq
\lambda_{Z,s2t,(5,6)} = \frac{1}{2}\Bigl [ 6v^2+4qv+q^2+4v^3+qv^2+v^4 \pm 
\sqrt{R_{256}} \ \Bigr ]
\label{lamtor2_56}
\eeq
where
\beq
R_{256}= (v^4+6v^3+8v^2+3qv^2+6qv+q^2)(v^4+2v^3+4v^2-qv^2+2qv+q^2) \ .
\label{r256}
\eeq
We note that $\lambda_{Z,s2t,3}\lambda_{Z,s2t,4}=v^3(v+1)^2(v+q)$ while 
$\lambda_{Z,s2t,5}\lambda_{Z,s2t,6}=v^2(v+1)^2(q+v)^2$. 
The corresponding coefficients for the torus and Klein bottle strips are 
\beq
c_{Z,s2t,1}=c^{(2)}
\label{c1tor2}
\eeq
\beq
c_{Z,s2t,j}=c^{(1)} \quad {\rm for} \ \ j=2,3,4
\label{c234tor2}
\eeq
\beq
c_{Z,s2t,j}=1 \quad {\rm for} \ \ j=5,6
\label{c56tor2}
\eeq
\beq
c_{Z,s2k,1}=-1
\label{c1kb2}
\eeq
\beq
c_{Z,s2k,2}=-c^{(1)}
\label{c2kb2}
\eeq
\beq
c_{Z,s2k,j}=c^{(1)} \quad {\rm for} \ \ j=3,4
\label{c34kb2}
\eeq
\beq
c_{Z,s2k,j}=1 \quad {\rm for} \ \ j=5,6 \ .
\label{c56kb2}
\eeq

 From these results we calculate the determinants
\beqs
D_Z(sq,2 \times L_x,PBC_y,PBC_x) & = & 
\prod_{j=1}^6 (\lambda_{Z,s2t,j})^{c_{Z,s2t,j}} \cr\cr
& = & 
v^{2q(q-1)}(v+1)^{2q}(v+q)^{2q} \cr\cr
& = & (y-1)^{2q^2}(yx)^{2q}
\label{lamprodtor2}
\eeqs
and
\beqs
D_Z(sq,2 \times L_x,PBC_y,TPBC_x) & = & 
\prod_{j=1}^6 (\lambda_{Z,s2t,j})^{c_{Z,s2k,j}} \cr\cr
& = & v^{2(q-1)}(v+1)^{2q}(v+q)^2 \cr\cr
& = & [(y-1)y]^{2q}x^2 \ . 
\label{lamprodkb2}
\eeqs

In the special case of the zero-temperature Potts antiferromagnet, $v=-1$, we
have $N_{P,s2t,\lambda}=N_{P,s2k,\lambda}=4$ and the corresponding reductions
\beq
\lambda_{Z,s2t,1} \rightarrow 1
\label{lam1tor2a0}
\eeq
\beq
\lambda_{Z,s2t,2} \rightarrow 1-q
\label{lam2tor2a0}
\eeq
\beq
\lambda_{Z,s2t,3} \rightarrow 3-q
\label{lam4tor2a0}
\eeq
\beq
\lambda_{Z,s2t,5} \rightarrow D_4 = q^2-3q+3
\label{lam5tor2a0}
\eeq
and
\beq
\lambda_{Z,s2t,j} \rightarrow 0 \quad {\rm for} \ \ j=4,6 \ .
\label{lam46tor2a0}
\eeq
In this limit the resultant partition functions (chromatic polynomials) for 
the cases of torus and Klein bottle boundary conditions are identical to 
those for cyclic and M\"obius boundary conditions, respectively.  This is a
special case of a general result that the chromatic polynomial for a graph $G$
and a graph $H$ which differs from $G$ in having multiple edges between given
sets of vertices are equal.

\subsection{$L_y=3$} 

We find
\beq
N_{Z,s3t,\lambda}=20
\label{nztottorus3}
\eeq
\beq
N_{Z,s3k,\lambda}=12 \ . 
\label{nztotklein}
\eeq
Recall that for the special case $v=-1$ of the zero-temperature Potts
antiferromagnet, \cite{tk}
\beq
N_{P,s3t,\lambda}=8
\label{nptottorus}
\eeq
\beq
N_{P,s3k,\lambda}=5 \ . 
\label{nptotklein}
\eeq
We calculate  
\beq
Z(sq,3 \times m,PBC_y,PBC_x,q,v) =
\sum_{j=1}^{20} c_{Z,s3t,j}(\lambda_{Z,s3t,j}(q,v))^m
\label{zgsums3torus}
\eeq
and
\beq
Z(sq,3 \times m,PBC_y,TPBC_x,q,v) =
\sum_{j=1}^{12} c_{Z,s3k,j}(\lambda_{Z,s3k,j}(q,v))^m
\label{zgsums3klein}
\eeq
where (ordering the $\lambda_{Z,s3t,j}$'s by decreasing degree of the 
associated coefficients $c_{Z,s3t,j}$)
\beq
\lambda_{Z,s3t,j}=\lambda_{Z,s3c,j} \quad {\rm for} \ \ 1 \le j \le 6
\label{lamtor1to6}
\eeq
\beq
\lambda_{Z,s3t,(7,8)}=\frac{v^2}{2}\biggl [ q+v(v+6) \pm \Bigl ( (q+v(v+6))^2
-4v(v+q)(v+1) \Bigr )^{1/2} \biggr ]
\label{lamtor78}
\eeq
\beq
\lambda_{Z,s3t,9}=v^3(v+1)
\label{lamtor9}
\eeq
\beq
\lambda_{Z,s3t,10}=v^2(v+q)(v+1) \ . 
\label{lamtor10}
\eeq
The next three terms coincide with three for the cyclic/M\"obius strip:
\beq
\lambda_{Z,s3t,j}=\lambda_{Z,s3c,j-4} \quad {\rm for} \ \ j=11,12,13 \ . 
\label{lamtor14to16}
\eeq
The $\lambda_{Z,s3t,j}$ for $14 \le j \le 17$ are roots of the quartic
\beq
\xi^4+g_{41}\xi^3+g_{42}\xi^2+g_{43}\xi+g_{44}=0
\label{eqquartor}
\eeq
where
\beq
g_{41}= -v(25v^2+17v^3+v^2q+8qv+v^5+6v^4+q^2)
\label{g41}
\eeq
\beqs
& & g_{42}=v^3(v+1)(q^3+48v^3+v^5q+42v^2q+2v^6+42v^4+30qv^3+5q^2v^2 \cr\cr
& & +12q^2v+10v^4q+16v^5+v^3q^2)
\label{g42}
\eeqs
\beqs
& & g_{43}=-v^6(v+1)^2(v+q)(2q^2+28v^3+25v^2+12qv+11v^2q+11v^4+4qv^3 \cr\cr
& & +q^2v^2+2q^2v+v^5)
\label{g43}
\eeqs
\beq
g_{44}=v^9(v+q)^3(v+1)^5 \ . 
\label{g44}
\eeq
Finally, $\lambda_{Z,s3t,j}$ for $j=18,19,20$ are roots of the cubic equation
\beq
\xi^3+g_{31}\xi^2+g_{32}\xi+g_{33}=0
\label{eqcub2}
\eeq
where
\beq
g_{31}=-(24v^3+16v^2q+16v^4+6q^2v+v^6+6v^5+q^3+2qv^3)
\label{h31}
\eeq
\beqs
& & g_{32}=v^2(v+1)(v+q)(q^3+24v^3+26v^2q+v^6+26v^4+20qv^3+5q^2v^2 \cr\cr
& & +10q^2v+5v^4q+10v^5+v^3q^2)
\label{h32}
\eeqs
\beq
g_{33}=-v^5(v+q)^4(v+1)^4 \ . 
\label{h33}
\eeq

The terms that enter in eq. (\ref{zgsums3klein}) are
\beq
\lambda_{Z,s3k,j}=\lambda_{Z,s3t,j} \quad {\rm for} \ \ j=1,2
\label{lamklein12}
\eeq
\beq
\lambda_{Z,s3k,j}=\lambda_{Z,s3t,j+4}  \quad {\rm for} \ \ 3 \le j \le 5
\label{lamklein3to5}
\eeq
\beq
\lambda_{Z,s3k,j}=\lambda_{Z,s3t,j+8} \quad {\rm for} \ \ 6 \le j \le 12 \ . 
\label{lamklein6to12}
\eeq
The eight terms in the Potts model partition function for the torus strip 
$\lambda_{Z,s3t,j}$ with $3 \le j \le 6$ and $10 \le j \le 13$ do not
occur for the Klein bottle strip.  From our calculations of $Z(G,q,v)$ for the
$L_y=2$ and $L_y=3$ strips of the square lattice with torus and Klein bottle
boundary conditions, we observe that the first two terms are consistent with
the same generalizations (\ref{lam1general}) and (\ref{lam2general}) as for the
cyclic and M\"obius strips, although the coefficients are, in general,
different. 

For the strip with torus boundary conditions the corresponding coefficients are
\beq
c_{Z,s3t,1}=q^3-6q^2+8q-1
\label{c1tor}
\eeq
\beqs
c_{Z,s3t,2} & = & c_{Z,s3t,9} = \frac{1}{2}c_{Z,s3t,3} = \frac{1}{2}c_{Z,s3t,4}
\cr\cr & = & \frac{1}{2}(c^{(2)}+c^{(0)})=\frac{1}{2}(q-1)(q-2) 
\label{c2349tor}
\eeqs
\beqs
c_{Z,s3t,j}=c^{(2)}-c^{(0)}=q(q-3) \quad {\rm for} 
\ \ j=5,6
\label{c56tor}
\eeqs
\beq
c_{Z,s3t,j}=\frac{1}{2}q(q-3) \quad {\rm for} \ \ j=7,8
\label{c78tor}
\eeq
\beq
c_{Z,s3t,10}=2
\label{c10tor}
\eeq
\beq
c_{Z,s3t,j}=2c^{(1)}=2(q-1) \quad {\rm for} \ \ 11 \le j \le 13
\label{c11to13tor}
\eeq
\beq
c_{Z,s3t,j}=q-1 \quad {\rm for} \ \ 14 \le j \le 17
\label{c14to17}
\eeq
\beq
c_{Z,s3t,j}=1 \quad {\rm for} \ \ 18 \le j \le 20 \ .
\label{c18to20tor}
\eeq
Note that $c_{Z,s3t,1}$ is the $d=3$ member of the sequence \cite{matmeth}
\beq
\kappa^{(d)}=\sum_{r=0}^d (-1)^{d-r}{d \choose r} q^{(r)} \ . 
\label{kappa}
\eeq

For the strip with Klein bottle boundary conditions the corresponding 
coefficients are
\beq
c_{Z,s3k,1}=-(q-1)
\label{c1klein}
\eeq
\beq
c_{Z,s3k,2}=-c_{Z,s3k,5}=-\frac{1}{2}(q-1)(q-2)
\label{c25klein}
\eeq
\beq
c_{Z,s3k,3}=c_{Z,s3k,4}=\frac{1}{2}q(q-3)
\label{c34klein}
\eeq
\beq
c_{Z,s3k,j}=c^{(1)}=q-1 \quad {\rm for} \ \ 6 \le j \le 9
\label{c6to9klein}
\eeq
\beq
c_{Z,s3k,j}=1 \quad {\rm for} \ \ 10 \le j \le 12 \ .
\label{c10to12klein}
\eeq

Using these results, we find that 
\beqs
D_Z(sq,3 \times L_x,PBC_y,PBC_x) & = & 
v^{3q^2(q-1)}(v+1)^{3q^2}(v+q)^{3q^2} \cr\cr
     & = & (y-1)^{3q^3}(yx)^{3q^2}
\label{lamprodtor3}
\eeqs
This and the analogous $L_y=2$ result, eq. (\ref{lamprodtor2}), can both be fit
by the formula
\beq
D_Z(sq,L_y \times L_x,PBC_y,PBC_x)= (y-1)^{L_y q^{L_y}}(yx)^{L_y q^{L_y-1}}
\ . 
\label{lamprodtorly}
\eeq
For the $L_y=3$ strip with Klein bottle boundary conditions, 
\beqs
D_Z(sq,3 \times L_x,PBC_y,TPBC_x) & = & v^{3q(q-1)}(v+1)^{q(q+2)}(v+q)^{3q}
\cr\cr & = & (y-1)^{3q^2}y^{q(q+2)}x^{3q}
\label{lamprodkb3}
\eeqs

\subsection{$v=-1$ Special Case}

For the special case $v=-1$, ($T=0$ Potts antiferromagnet), the 20 (12) terms 
in the respective partition functions for the $L_y=3$ strip of the square
lattice with torus (Klein bottle) boundary conditions reduce to 8 (5) terms
entering in the chromatic polynomials for these strips.  We have 
\beq
P(sq,3 \times m,PBC_y,PBC_x,q)\sum_{j=1}^8 c_{P,s3t,j}(\lambda_{P,s3t,j})^m
\label{psumtor}
\eeq
\beq
P(sq,3 \times m,PBC_y,TPBC_x,q)=\sum_{j=1}^5 c_{P,s3k,j}(\lambda_{P,s3k,j})^m
 \ .
\label{psumklein}
\eeq
We order the terms $\lambda_{P,s3t,j}$ by decreasing degree of the
associated coefficients $c_{P,s3t,j}$ for the torus strip, as polynomials in
$q$ (this differs from the order and hence labelling used in
\cite{tk}). These terms are 
\beq 
\lambda_{P,s3t,1}=-1
\label{lamptor1}
\eeq
\beq
\lambda_{P,s3t,2}=q-1
\label{lamptor2}
\eeq
\beq
\lambda_{P,s3t,3}=q-4
\label{lamptor3}
\eeq
\beq
\lambda_{P,s3t,4}=q-2
\label{lamptor4}
\eeq
and
\beq
\lambda_{P,s3t,5}=q-5
\label{lamptor5}
\eeq
\beq
\lambda_{P,s3t,6}=-(q^2-7q+13)
\label{lamptor6}
\eeq
\beq
\lambda_{P,s3t,7}=-(q-2)^2
\label{lamptor7}
\eeq
\beq
\lambda_{P,s3t,8}=q^3-6q^2+14q-13 \ .
\label{lamptor8}
\eeq
Thus, the correspondences for $v=-1$ are as follows:
\beq
\lambda_{Z,s3t,j} \rightarrow \lambda_{P,s3t,j} \quad {\rm for} \ \ j=1,2,3
\label{lam13tora0}
\eeq
\beq
\lambda_{Z,s3t,5} \rightarrow \lambda_{P,s3t,4}
\label{lam5tora0}
\eeq
\beq
\lambda_{Z,s3t,7} \rightarrow \lambda_{P,s3t,5}
\label{lam7tora0}
\eeq
\beq
\lambda_{Z,s3t,11} \rightarrow \lambda_{P,s3t,7}
\label{lam11tora0}
\eeq
\beq
\lambda_{Z,s3t,14} \rightarrow \lambda_{P,s3t,6}
\label{lam14tora0}
\eeq
\beq
\lambda_{Z,s3t,18} \rightarrow \lambda_{P,s3t,8}
\label{lam18tora0}
\eeq
\beq
\lambda_{Z,s3t,j} \rightarrow 0 \quad {\rm for} \ \ 
j=4,6,8,9,10,12,13,15,16,17,19,20 \ .
\label{lamjtora0}
\eeq 
The reductions for the $\lambda_{Z,s3k,j}$ when $v=-1$ follow from these
and the relations (\ref{lamklein12})-(\ref{lamklein6to12}). 

\subsection{$v=0$ Special Case}

For $v=0$, i.e., infinite temperature, we have
\beq
\lambda_{Z,s3t,18} = q^3
\label{lam18torv0}
\eeq
and
\beq
\lambda_{Z,s3t,j} = 0 \quad {\rm for} \ \ j \ne 18
\label{lamjtorv0}
\eeq
so that $Z(sq,3 \times m,PBC_y,(T)PBC_x,q,v=0)=q^n$.

\section{Thermodynamic Properties of the Potts Model on 
Strips of the Square Lattice}

\subsection{General} 

We can use our exact results to study a number of interesting physical
thermodynamic properties.  Some of our results apply for arbitrarily great
width $L_y$ while others involve comparisons of specific widths for which we
have carried out exact calculations. The thermodynamic properties are
independent of the longitudinal boundary conditions (i.e., in the
infinite-length direction) but do depend on the transverse boundary conditions
(finite-width direction).  For the strips, for any $L_y$ no matter how large,
the ferromagnet is critical only at $T=0$, and as $T \to 0$ and the correlation
length $\xi \to \infty$, the strip acts as a one-dimensional system, since
$\lim_{L_x \to \infty} L_y/L_x=0$.  In contrast, for the Potts ferromagnet on
the square lattice, the phase transition occurs at finite temperature, at the
known value $K_c = \ln(1+\sqrt{q} \ )$. Thus, studies of the thermodynamic
behavior of the Potts model for general $q$ on $L_y \times \infty$ strips 
complement studies such as those on the approach to the thermodynamic limit of
the Ising model on $L_x \times L_y$ rectangular regions, in which $L_x$ and
$L_y$ both get large with a fixed finite ratio $L_y/L_x$ \cite{fisherferd}, and
finite-size scaling analyses \cite{fisherbarber,barber}.  

For the infinite-length limit of the $L_y=3$
strip with free transverse boundary conditions (i.e., the cyclic or M\"obius
strips considered here and also the free strip), the reduced free energy per 
site is given by 
\beq 
f = \frac{1}{3}\ln \lambda_{Z,s3c,17}
\label{fcyclic3}
\eeq
while for the infinite-length limits of the strips with periodic
transverse boundary conditions (i.e. the torus and Klein bottle strips
considered here as well as the cylindrical strip), we have, for $L_y=2$, 
\beq
f = \frac{1}{2}\ln \lambda_{Z,s2t,5}
\label{ftorus2}
\eeq
and, for $L_y=3$, 
\beq
f = \frac{1}{3}\ln \lambda_{Z,s3t,18} \ . 
\label{ftorus3}
\eeq 
It is straightforward to calculate from eqs. (\ref{fcyclic3})-(\ref{ftorus3}) 
the respective expressions for the internal energy $U$ and specific heat $C$ 
per site.  In all cases, as noted above, 
the Potts ferromagnet has a $T=0$ critical point for
general $q$.  The Potts antiferromagnet has a $T=0$ critical point if $q$ has a
certain value or values depending on the given strip; for the $L_y=2$ strip
with free or periodic transverse boundary conditions, this value is $q=2$ 
\cite{w}, while for the $L_y=3$ strip with (i) free and (ii) 
periodic transverse boundary conditions, the antiferromagnet has $T=0$ 
critical points at (i) $q=2$ and $q \simeq 2.34$ \cite{wcyl,wcy} (see eq. 
(\ref{qccyclly3}) below), and (ii) $q=2$ and $q=3$, respectively \cite{tk}. 

 For the Ising value $q=2$, since the $L_x \to \infty$ limit can be taken with
even $L_x$ for which the cyclic strip is bipartite, the ferromagnet and
antiferromagnet are equivalent, and hence have equivalent $T=0$ critical
points.  However, in the antiferromagnetic case, because of the
noncommutativity (\ref{fnoncomm}), to avoid pathologies of the type discussed
in detail in \cite{a}, one should set $q=2$ first and then take the $n \to
\infty$ limit, i.e. one must use $f_{nq}$ rather than $f_{qn}$.  In contrast,
the $L_y=3$ strip with periodic transverse boundary conditions is not bipartite
and in the $q=2$ Ising case, antiferromagnetic ordering involves frustration.
Nevertheless, the model still has a $T=0$ critical point, as is clear from the
fact that the singular locus ${\cal B}_q$ runs through the point $q=2$ in the
$q$ plane.  This is somewhat analogous to the 2D Ising antiferromagnet on the
triangular lattice in the sense that both involve zero-temperature critical
points with frustration.

\subsection{Potts Ferromagnet}

\subsubsection{Free Energy, Specific Heat, and Approach to Square Lattice
Limit} 

\begin{figure}[hbtp]
\centering
\leavevmode
\epsfxsize=5in
\begin{center}
\leavevmode
\epsffile{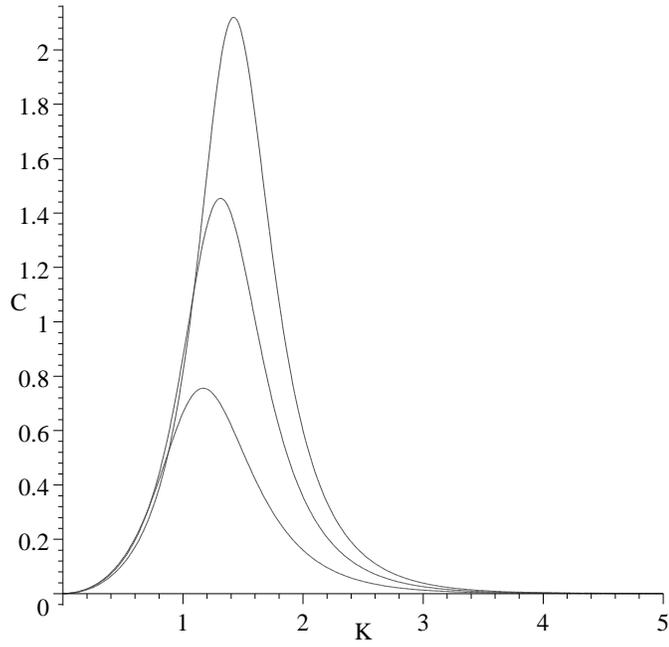}
\end{center}
\vspace{-25mm}
\caption{\footnotesize{Specific heat (with $k_B \equiv 1$) for the Potts
ferromagnet on the infinite-length, width $L_y=3$ strip of the square lattice
with free transverse boundary conditions, as a function of $K=J/(k_BT)$.  Going
from bottom to top in order of the heights of the maxima, the curves are for
$q=2,3,4$.}}
\label{cfmsqpxy3strip}
\end{figure}

\begin{figure}[hbtp]
\centering
\leavevmode
\epsfxsize=5in
\begin{center}
\leavevmode
\epsffile{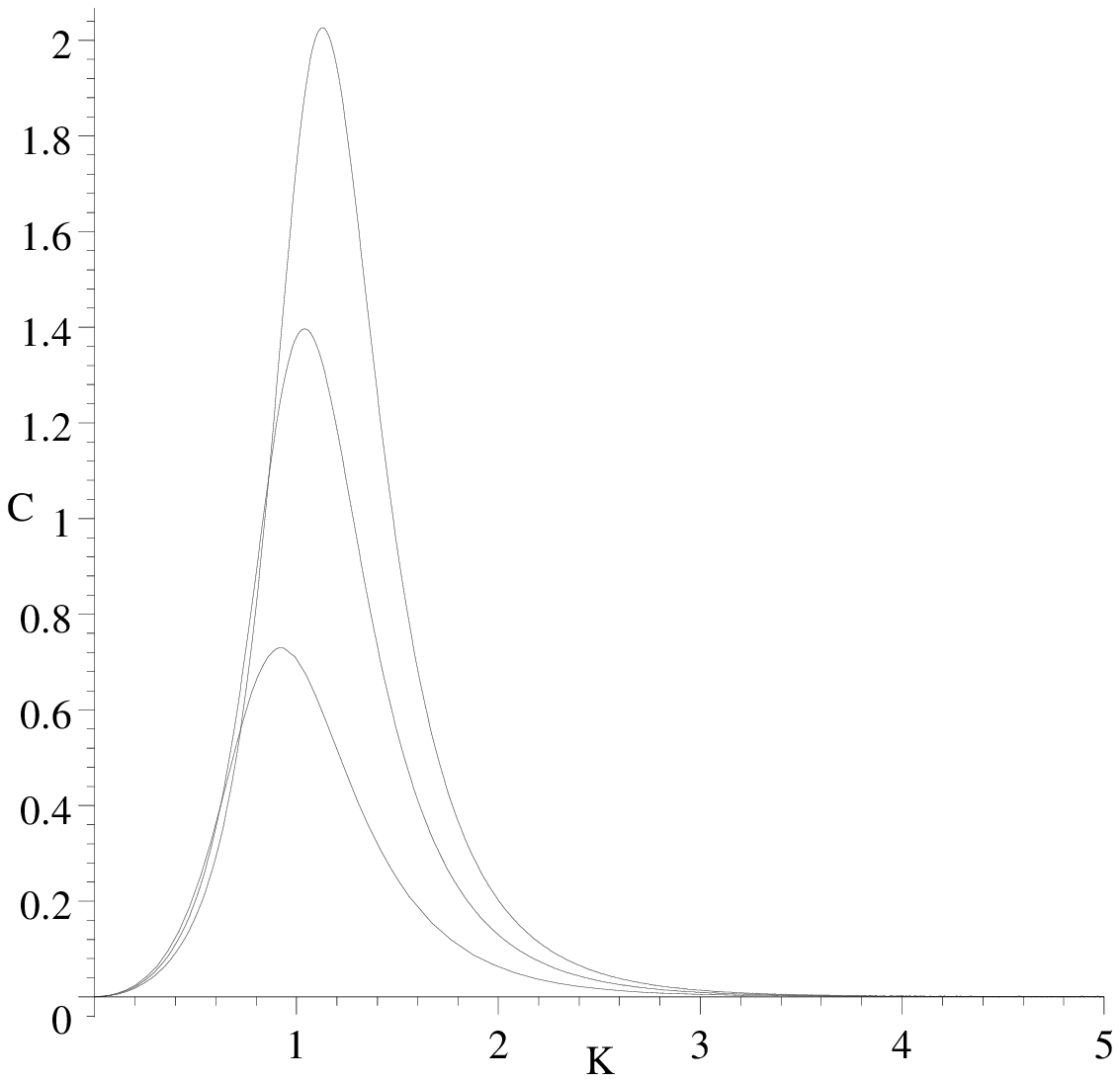}
\end{center}
\vspace{-25mm}
\caption{\footnotesize{Specific heat (with $k_B \equiv 1$) for the Potts
ferromagnet on the infinite-length, width $L_y=3$ strip of the square lattice
with periodic transverse boundary conditions as a function of $K=J/(k_BT)$.
Going from bottom to top in order of the heights of the maxima, the curves are
for $q=2,3,4$.}}
\label{cfmsqpxpy3strip}
\end{figure}

\begin{table}
\caption{\footnotesize{Some properties of the specific heat of the $q$-state
Potts ferromagnet on strips of the square lattice with free and periodic
transverse boundary conditions.}}
\begin{center}
\begin{tabular}{|c|c|c|c|c|c|}
\hline\hline $q$ & $BC_y$ & $L_y$ & $K_{C,max}$ & $C_{max}/k_B$ &
$T_{C,max}/T_{c,sq}$ \\ \hline\hline
2 & $FBC_y$ & 1 & 2.40 & 0.439 & 0.367 \\ \hline
2 & $FBC_y$ & 2 & 1.33 & 0.631 & 0.663 \\ \hline
2 & $FBC_y$ & 3 & 1.17 & 0.755 & 0.753 \\ \hline\hline
2 & $PBC_y$ & 2 & 1.00 & 0.529 & 0.8805 \\ \hline
2 & $PBC_y$ & 3 & 0.92 & 0.729 & 0.958 \\ \hline\hline
3 & $FBC_y$ & 1 & 2.65 & 0.762 & 0.379 \\ \hline
3 & $FBC_y$ & 2 & 1.49 & 1.167 & 0.676 \\ \hline
3 & $FBC_y$ & 3 & 1.31 & 1.453 & 0.767 \\ \hline\hline
3 & $PBC_y$ & 2 & 1.12 & 0.943 & 0.899 \\ \hline
3 & $PBC_y$ & 3 & 1.04 & 1.396 & 0.966 \\ \hline\hline
4 & $FBC_y$ & 1 & 2.84 & 1.023 & 0.386 \\ \hline
4 & $FBC_y$ & 2 & 1.61 & 1.647 & 0.684 \\ \hline
4 & $FBC_y$ & 3 & 1.42 & 2.119 & 0.774 \\ \hline\hline
4 & $PBC_y$ & 2 & 1.21 & 1.292 & 0.910 \\ \hline
4 & $PBC_y$ & 3 & 1.13 & 2.026 & 0.972 \\ \hline\hline
\end{tabular}
\end{center}
\label{cfmtable}
\end{table}

In Figs. \ref{cfmsqpxy3strip} and \ref{cfmsqpxpy3strip} we show plots of the
specific heat for the Potts ferromagnet on the infinite-length limits of the
$L_y=3$ strips with free and periodic transverse boundary conditions.  In each
curve, the specific heat assumes a maximum value $C_{max}$ at a certain value
of inverse temperature $K_{C,max}$, or equivalently, temperature 
$T_{C,max}=J/(k_B K_{C,max})$.  It is of interest to inquire how these
quantities depend on $q$ and the transverse boundary conditions.  We give this
information for the results in the present paper and also our relevant 
previous studies \cite{a} in Table \ref{cfmtable}.  
In general, for the Potts ferromagnet on the infinite-length strip of width
$L_y$ and given transverse boundary condition, we find that, for physical $q$, 
\beq
T_{C,max} \ \ {\rm is \ \ a \ \ decreasing \ \ function \ \ of} \ \ q
\label{tcmaxqinc}
\eeq
and
\beq
C_{max} \ \ {\rm is \ \ an \ \ increasing \ \ function \ \ of} \ \ q \ . 
\label{cmaxqinc}
\eeq
The monotonicity (\ref{tcmaxqinc})
 can be understood as a consequence of the fact that the
specific heat, $C=dU/dT$, reaches a maximum at the temperature where there is
an onset of short-range order, as reflected in the maximal temperature
derivative of the internal energy.   As $q$ increases, there are more effective
degrees of freedom per site, and hence it is necessary to cool the system to a
lower temperature, i.e., higher value of $K$, to obtain short-range ordering. 
The monotonicity relation (\ref{cmaxqinc}) shows that this onset of short-range
order occurs more sharply as $q$ increases. 

We may also study the dependence of $T_{C,max}$ and $C_{max}$ as a function of
the strip width and transverse boundary conditions for the Potts ferromagnet
with a given value of $q$ and investigate how $T_{C,max}$ approaches the
critical value \cite{wurev}
\beq
T_c = \frac{J}{k_B \ln (1+\sqrt{q})} \ , 
\label{tcfm}
\eeq
i.e., $a_c=1+\sqrt{q}$, of the model on the infinite square lattice.  
As summarized in Table \ref{cfmtable}, we find that, for a 
given (physical, integral) $q \ge 2$ and for strips considered here, 
\beq
T_{C,max} \ \ {\rm and} \ \ C_{max} \ \ {\rm are \ \ increasing \ \ functions 
\ \ of} \ \ L_y \ . 
\label{tcmaxlyinc}
\eeq 
For the strips with free transverse boundary conditions, the monotonicity
relation (\ref{tcmaxlyinc}) for $T_{C,max}$ can be explained as follows.  As
$L_y$ increases, the effective coordination number, given by (\ref{deltaeff})
above, also increases.  Since the short-range ordering is a consequence of the
ferromagnetic spin-spin interactions and since, on average, a given spin feels
the influence of $\Delta_{eff}$ of these interactions, a given degree of
short-range order can be established at a higher temperature as $L_y$
increases.  Since, as recalled above, the maximum in the specific heat occurs
at the temperature where this onset of short range order is taking place, the
monotonicity relation (\ref{tcmaxlyinc}) follows.  Of course, in the limit $L_y
\to \infty$, the point $T_{C,max}$ is a true critical point at which there is a
nonanalyticity in the free energy, and the specific heat does not just reach a
maximum due to short-range ordering, but diverges (logarithmically) \cite{ons},
with the onset of long-range order, i.e.  nonzero spontaneous
magnetization. For the strips with periodic transverse boundary conditions, as
$L_y$ increases, there is no change in the coordination number $\Delta=4$;
rather, there are actually two countervailing tendencies: (i) an increase in
the effectiveness of short-range ordering, which, in the limit $L_y \to \infty$
yields long-range order at sufficiently low temperature, (ii) a decrease in the
finite-size effects involving spin-spin interactions that loop around the
finite transverse periodic direction.  To explain (ii), consider two spins
which, for simplicity, are located at the same value of $x$ but different
values of $y$, $y_1$ and $y_2$.  On an infinite square lattice, the leading
contribution to the interaction of these spins would be via a minimum-distance
path, of length $r=|y_1-y_2|$.  However, for finite $L_y$, there is another
contribution, namely via the path going the other way around the transverse
direction, of length $L_y-r$.  This is a finite-lattice artifact, the effect of
which goes to zero as $L_y \to \infty$.  For the toroidal strips considered
here, the effect (i) outweighs (ii), yielding the monotonicity relation
(\ref{tcmaxlyinc}).  This result may be compared with the results for the Ising
$q=2$ case obtained by Ferdinand and Fisher \cite{fisherferd}.  These authors
considered finite-size $L_x \times L_y$ lattices with toroidal boundary
conditions and studied the limiting behavior of the specific heat as $L_x \to
\infty$ and $L_y \to \infty$ with fixed finite nonzero ratio $L_y/L_x$.  Among
their other results, they showed that if this limit is taken with finite-size
sections with relatively square-like aspect ratios $1/3.1 \lsim L_y/L_x \lsim
3.1$, then $T_{C,max}$ approaches $T_c$ from above in the thermodynamic limit,
while for more narrow rectangular shapes, $L_y/L_x \gsim 3.1$ or $L_y/L_x <
1/3.1$, $T_{C,max}$ approaches $T_c$ from below.  Although our finite-size
studies of the $q=2$ Potts model on toroidal strips do not, strictly speaking,
fall into the category considered by Ferdinand and Fisher since we take $L_x
\to \infty$ keeping $L_y$ fixed, so that $L_y/L_x$ is identically zero, our
results may be expected to correspond most closely to the extreme limit
$L_y/L_x \to 0$ in the work of \cite{fisherferd}. Hence, one would expect that
for $q=2$ our $T_{C,max}$ should approach $T_c$ from below, and it does.  It is
of interest that our calculations generalize this monotonicity result to 
$q \ge 3$. 

Our calculations also show that for a $q$ and given width $L_y$, the ratio 
$T_{C,max}/T_c$ is closer to its limit of unity if one uses a strip with
periodic rather than free transverse boundary conditions.  This is easily
explained since periodic transverse boundary conditions remove edge effects and
thereby reduce finite-size artifacts.

Although the (zero-field) free energy of the general $q$-state Potts model has
not been calculated for arbitrary temperature, it has been calculated in
\cite{baxtercrit} at the phase transition temperature $T_c$ for the 
ferromagnet, (\ref{tcfm}).  Of course, the free energy is also known exactly
for the $q=2$ Ising case.  In addition to the studies of $L_y$-dependence and
comparisons of $T_{C,max}/T_c$ discussed in the preceding paragraphs, it is of
interest to compare our results with a conformal field theory relation
concerning finite-size effects, for the $q$ values where the Potts ferromagnet
has a second-order transition in 2D, namely, $q=2,3$, and 4.  For this 
purpose, we recall that conformal field theory methods have provided insight 
into the universality classes of continuous, second-order phase transitions
and the associated critical exponents in terms of Virasoro
algebras with given central charges and scaling dimensions
\cite{isz}-\cite{dfms}.  The
Virasoro algebra with central extension depending on the central charge $c$ is
\beq
[L_m, L_n] = (m-n)L_{m+n} + \frac{c}{12}m(m^2-1)\delta_{m+n,0}
\label{virasoro}
\eeq
For the values of $q$ where the 2D Potts ferromagnet has a continuous 
transition (with infinite correlation length), the central charges are given 
by \cite{isz}-\cite{dfms} (i) $c=1/2$ for $q=2$,
(ii) $c=4/5$ for $q=3$, and (iii) $c=1$ for $q=4$. 
There is a useful relation from conformal field theory that describes the
approach of the free energy of an infinite strip with free or periodic
transverse boundary conditions to the 2D thermodynamic limit as $L_y \to
\infty$ at the critical temperature of the 2D model 
\cite{cardy}-\cite{affleck}:
\beq
f_{strip,L_y} = f_{bulk} + \frac{f_{surf.}}{L_y} + \frac{\Delta}{L_y^2} +
O(L_y^{-3})
\label{frel}
\eeq
where $f_{surf.}=0$ is nonzero (zero) for free (periodic) boundary
conditions in the $y$ direction and
\beq
\Delta = \left \{ \begin{array}{ll}
      \frac{\pi}{6}c & \mbox{for PBC$_y$} \\
     \frac{\pi}{24}c & \mbox{for FBC$_y$}
     \end{array} \right .
\label{delta}
\eeq
To compare our results with this relation, we use as input the critical values
of the Potts model free energy on the square lattice.  For $q=2$, from the
Onsager solution \cite{ons} one has
\beq
f_{Potts,q=2,crit.}=\frac{1}{2}\ln 2 + \ln(1+\sqrt{2} \ ) + \frac{2}{\pi}G_c = 
1.811068985...
\label{fcpottsq2}
\eeq
where $G_C$ is the Catalan constant
\beq
G_C = \sum_{k=0}^{\infty}\frac{(-1)^k}{(2k+1)^2}= 0.91596559...
\label{gc}
\eeq
For $q=3$ and $q=4$, one has \cite{baxtercrit}
\beq
f_{Potts,q=3,crit.}=\ln 2 + \frac{1}{2}\ln 3 + \frac{1}{3}\ln(2+\sqrt{3} \ ) + 
\frac{4G_C}{3\pi} = 2.07018716...
\label{fcpottsq3}
\eeq
and
\beq
f_{Potts,q=4,crit.}= -3\ln 2 + 
4\ln \biggl ( \frac{\Gamma(1/4)}{\Gamma(3/4)} \biggr ) = 2.25952475... 
\label{fcpottsq4}
\eeq
where $\Gamma(z)$ is the Euler Gamma function. 
As an illustration, we consider the width $L_y=3$ strip with periodic
transverse boundary conditions.  In Table \ref{cfttable} we list the values of
the right-hand side of (\ref{frel}) and the evaluation of $f_{strip}$ at the
value $T=T_c$ where the model is critical on the 2D lattice.  Of course, for
any finite value of $L_y$, $f_{strip}$ is not critical at this value of
temperature and, for comparison, we also list the comparison with $f_{strip}$
evaluated at $T=T_{C,max}$.  Clearly, as $L_y \to \infty$, $T_{C,max} \to T_c$,
so that these are equivalent in this limit.  One sees that eq. (\ref{frel}) 
is reasonably well satisfied, even for the modest value of the width $L_y=3$.

\begin{table}
\caption{\footnotesize{Comparison of exact partition function calculations with
conformal field theory finite-size relation for the infinite-length, width 
$L_y=3$ strip with periodic transverse boundary conditions.  See text for 
further details.}}
\begin{center}
\begin{tabular}{|c|c|c|c|}
\hline\hline $q$ & RHS & $f_{strip}(T_c)$ & $f_{strip}(T_{C,max})$ \\ 
\hline\hline
2 & 1.840 & 1.8425 & 1.91  \\ \hline
3 & 2.117 & 2.121  & 2.18  \\ \hline
4 & 2.318 & 2.324  & 2.37  \\ \hline
\end{tabular}
\end{center}
\label{cfttable}
\end{table}

\subsubsection{$T \to 0$ Limit}

We next derive some general results concerning the low-temperature behavior of
the (reduced) free energy $f$ and, from this, the internal energy $U$ and
specific heat $C$, for the $q$-state Potts ferromagnet on an infinite-length
strip of the square lattice with arbitrary longitudinal boundary conditions and
specified transverse boundary conditions.  For technical convenience, we use
free longitudinal boundary conditions for our calculations, assume a large
length $L_x$ when calculating the partition function, and then, as usual, take
$L_x \to \infty$ when calculating the free energy (per site). At $T=0$, $Z=q
a^{e(G)}$, where, as defined above, $e(G)$ denotes the number of edges (bonds)
on the strip graph.  Now consider finite-temperature corrections to this
result.  If the lattice were of dimensionality $d \ge 2$, the leading order
term always arises from a configuration in which one changes one spin from its
preferred value to one of the $(q-1)$ other values, and so forth for successive
corrections.  However, the leading corrections do not, in general, arise in
this simple a manner for the finite-width strips, especially in the case where
the transverse boundary conditions are free.  Let us begin with these.

For $L_y=1$, the leading low-temperature correction arises from changing all of
the spins to the right, say, of a given point from their previous value to one
of the $(q-1)$ other values.  Since this can be done at any of $L_x$ locations,
we obtain $f = K + (q-1)e^{-K} + ...$, where $...$ here and below indicate
subdominant terms in the $T \to 0$ limit, and hence $U=J[-1 + (q-1)e^{-K}]$ and
$C=k_BK^2(q-1)e^{-K}$.  Of course, in this $L_y=1$ case, the exact expressions
for $f$, $U$, and $C$ are simple: $f=\ln(e^K+q-1)$, $U=-Je^K/(e^K+q-1)$, and
$C=k_BK^2(q-1)e^K/(e^K+q-1)^2$.

For $L_y=2$, the
leading correction to the low-temperature limit for $Z$ is obtained by changing
all of the spins to the right of a given transverse slice from their original
value to one among the $(q-1)$ other possibilities.  This yields the expansion
\beq
f_{FBC_y,L_y=2} = \frac{3}{2}K +  \frac{1}{2}(q-1)e^{-2K} + ... 
\label{fcyc2}
\eeq
and hence 
\beq
U_{FBC_y,L_y=2} = J\Bigl [ -\frac{3}{2} + (q-1)e^{-2K} + ... \Bigr ] 
\label{ucyc2}
\eeq
and
\beq
C_{FBC_y,L_y=2} = 2k_BK^2(q-1)e^{-2K} + ...
\label{ccyc2}
\eeq
as given in \cite{a}, in agreement with a direct derivation from the exact
results obtained therein.  

For $L_y=3$, there are two different types of leading corrections to the 
low-temperature limit for $Z$, obtained by (i) changing
all of the spins to the right of a given transverse slice from their original
value to one among the $(q-1)$ other possibilities, and (ii) changing a spin on
the upper or lower boundary from its original value to one among the $(q-1)$
others.  These both involve a suppression factor $e^{-3K}$, and one gets the 
expansion
\beq
f_{FBC_y,L_y=3} = \frac{5}{3}K +  (q-1)e^{-3K} + ...
\label{fcyc3}
\eeq
and hence
\beq
U_{FBC_y,L_y=3} = J\Bigl [ -\frac{5}{3} + 3(q-1)e^{-3K} + ... \Bigr ]
\label{ucyc3}
\eeq
and
\beq
C_{FBC_y,L_y=3} = 9k_BK^2(q-1)e^{-3K} + ...
\label{ccyc3}
\eeq
We have that this is consistent with a direct derivation from the exact
result (\ref{fcyclic3}).

For greater widths, $L_y \ge 4$, the leading 
low-temperature correction arises from changing one of the spins on the
upper or lower horizontal edges from their original values to one of the other
$(q-1)$ possible values.  This makes the dominant contribution because of the
fact that these edge spins have a lower coordination number (degree), namely
$\Delta=3$, than the spins in the interior of the strip.  We have 
\beq
f_{FBC_y,L_y \ge 4} = \frac{\Delta_{eff}}{2}K + \frac{2}{L_y}(q-1)e^{-3K} + ...
\label{fcycge4}
\eeq
where $\Delta_{eff}$ was given in eq. (\ref{deltaeff}) above, and hence 
\beq
U_{FBC_y,L_y \ge 4} = J\Bigl [ -\frac{\Delta_{eff}}{2} + 
\frac{6}{L_y}(q-1)e^{-3K} + ... \Bigr ]
\label{ucycge4}
\eeq
and
\beq
C_{FBC_y,L_y \ge 4} = \frac{18}{L_y}k_BK^2(q-1)e^{-3K} + ...
\label{ccycge4}
\eeq
The essential zeros in $U$ and $C$ are
typical for a spin model at its lower critical dimensionality and mean that the
low-temperature series expansion has zero radius of convergence.  Physically,
this reflects the qualitatively different behavior at $T=0$, where there is
saturated magnetization, and at any nonzero temperature, regardless of how
small, where this magnetization vanishes identically.

Now consider a strip of the square lattice with free transverse boundary
conditions and arbitrarily great width $L_y \ge 3$.  Let the free energy be
denoted as 
\beq
f = \frac{1}{L_y}\ln \lambda_{L_y,PM}
\label{fgen}
\eeq
where $\lambda_{L_y,PM}$ is the term in (\ref{zgsum}) that is dominant for
physical temperature (paramagnetic phase).  Then for $L_y = 3$ this has 
the low-temperature expansion 
\beq
\lambda_{L_y=3,PM} = e^{5K}\Bigl [ 1 + 3(q-1)e^{-3K} + ... \Bigr ]
\label{lamlowcycly3}
\eeq
i.e., in the notation of eq. (\ref{lamu}),
\beq
\lambda_{L_y=3,PM,u}=1+3(q-1)u^3 + ...  \quad {\rm for} \ \ u \to 0
\label{lamugenly3}
\eeq
and for $L_y \ge 4$, 
\beq
\lambda_{L_y,PM} = e^{(\Delta_{eff}/2)L_yK}\Bigl [ 1 + 2(q-1)e^{-3K} + ... 
\Bigr ]
\label{lamlowcycge3}
\eeq
and hence, in the notation of eq. (\ref{lamu}),
\beq
\lambda_{L_y,PM,u}=1+2(q-1)u^3 + ...  \quad {\rm for} \ \ u \to 0 \ . 
\label{lamugen}
\eeq
Since the singular locus ${\cal B}$ is defined in the neighborhood of the 
point $u=0$ by the degeneracy of magnitudes
$|\lambda_{L_y,PM}|=|\lambda_{L_y,j}|$, where $\lambda_{L_y,j}$ is another
leading term at this point, it follows that this locus does not involve
an arbitrarily large number of curves crossing in an intersection point at
$u=0$, and the corresponding complex-temperature partition function zeros 
(Fisher zeros \cite{fisher}) do not become dense in the vicinity of the origin
as $L_y \to \infty$.  These results supercede a conjecture given in \cite{a}. 

We next derive corresponding low-temperature expansions for strips with 
periodic transverse boundary conditions.  In the cases $L_y=2$ and $L_y=3$, 
the 
leading correction term arises from changing all of the spins to the right of 
a transverse slice from their original values to one of $(q-1)$ other values.
This yields 
\beq
f_{PBC_y,L_y} = 2K + \frac{1}{L_y}(q-1)e^{-L_yK} + ...
\label{ftor2}
\eeq
\beq
U_{PBC_y,L_y} = J\Bigl [ -2 + (q-1)e^{-L_yK} + ... \Bigr ]
\label{utor2}
\eeq
\beq
C_{PBC_y,L_y} = L_yk_BK^2(q-1)e^{-L_yK} + ...
\label{ctor2}
\eeq
For $L_y=4$, these spin configurations involving a seam make a contribution
equal to those from configurations in which one changes one of the spins on a
transverse slice to one of $(q-1)$ other values, so that 
\beq
f_{PBC_y,L_y=4} = 2K + \frac{5}{4}(q-1)e^{-4K} + ...
\label{ftor4}
\eeq
\beq
U_{PBC_y,L_y=4} = J\Bigl [ -2 + 5(q-1)e^{-4K} + ... \Bigr ]
\label{utor4}
\eeq
\beq
C_{PBC_y,L_y=4} = 20k_BK^2(q-1)e^{-4K} + ...
\label{ctor4}
\eeq
Finally, for $L_y \ge 5$, the leading correction term in the low-temperature
limit arises in the same way as on an infinite 2D square lattice with toroidal
boundary conditions, namely via the change of a single spin to one of $(q-1)$
other values.  This yields expansions whose leading term is the same as on the
infinite square lattice:
\beq
f_{PBC_y,L_y \ge 5} = 2K + (q-1)e^{-4K} + ...
\label{ftorge5}
\eeq
\beq
U_{PBC_y,L_y \ge 5} = J\Bigl [ -2 + 4(q-1)e^{-4K} + ... \Bigr ]
\label{utorge5}
\eeq
\beq
C_{PBC_y,L_y \ge 5} = 16k_BK^2(q-1)e^{-4K} + ...
\label{ctorge5}
\eeq

Next consider a strip of the square lattice with periodic transverse boundary
conditions and arbitrarily great width $L_y \ge 5$.  Let the free energy be
denoted as in (\ref{fgen}).  Then we have the low-temperature expansion 
\beq
\lambda_{L_y,PM} = e^{2L_yK}\Bigl [ 1 + L_y(q-1)e^{-4K} + ...
\Bigr ]
\label{lamlowtorge5}
\eeq
i.e., in the notation of eq. (\ref{lamu}),
\beq
\lambda_{L_y,PM,u}=1+L_y(q-1)u^4 + ...  \quad {\rm for} \ \ u \to 0 \ . 
\label{lamutorgen}
\eeq

\subsection{Potts Antiferromagnet}

\subsubsection{Free Energy, Specific Heat, and Approach to Square Lattice
Limit}

 In the case of the Potts antiferromagnet, as discussed in \cite{a,ta}, there
can be nonanalyticities at sufficiently small non-integral $q$ and temperature,
but these do not represent physical phase transitions and involve a number of
pathologies such as non-existence of a thermodynamic limit independent of
longitudinal boundary conditions, negative specific heats, and lack of a Gibbs
measure.  For the physical cases, the analyticity of the free energy is a
consequence of the general theorem that one-dimensional and
quasi-one-dimensional spin systems with short-range spin-spin interactions do
not have a finite-temperature phase transition, which, in turn, is proved by an
elementary application of a Peierls argument.  This analyticity property is
equivalent to the property that the singular locus ${\cal B}_u$ does not cross
the positive real $u$ axis.

\begin{figure}[hbtp]
\centering
\leavevmode
\epsfxsize=5in
\begin{center}
\leavevmode
\epsffile{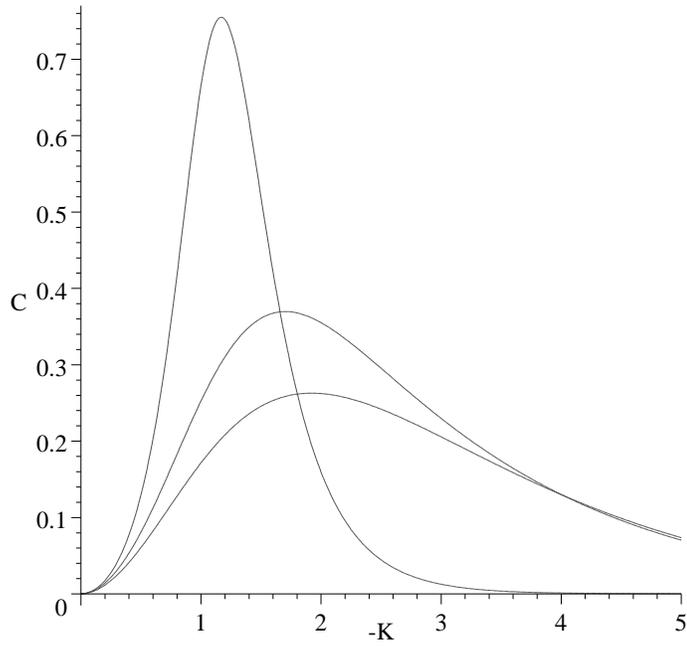}
\end{center}
\vspace{-25mm}
\caption{\footnotesize{Specific heat (with $k_B \equiv 1$) for the Potts
antiferromagnet on the infinite-length, width $L_y=3$ strip of the square
lattice with free transverse boundary conditions, as a function of
$K=J/(k_BT)$.  Going from top to bottom in order of the heights of the
maxima, the curves are for $q=2,3,4$.}}
\label{cafmsqpxy3strip}
\end{figure}

\begin{figure}[hbtp]
\centering
\leavevmode
\epsfxsize=5in
\begin{center}
\leavevmode
\epsffile{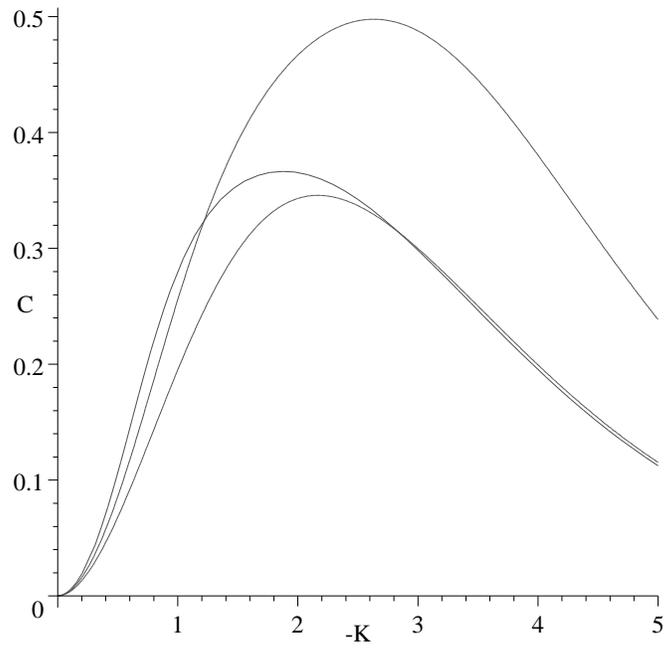}
\end{center}
\vspace{-25mm}
\caption{\footnotesize{Specific heat (with $k_B \equiv 1$) for the Potts
antiferromagnet on the infinite-length, width $L_y=3$ strip of the square 
lattice, as a function of $K=J/(k_BT)$.  Going from top to bottom in order of
the heights of the maxima, the curves are for $q=3,2,4$.}}
\label{cafmsqpxpy3strip}
\end{figure}

\begin{table}
\caption{\footnotesize{Some properties of the specific heat of the $q$-state
Potts antiferromagnet on strips of the square lattice with free and periodic
transverse boundary conditions.  Since for $q \ge 3$ this model has no
finite-temperature phase transition on the square lattice, the last column is
left blank.}}
\begin{center}
\begin{tabular}{|c|c|c|c|c|c|}
\hline\hline $q$ & $BC_y$ & $L_y$ & $-K_{C,max}$ & $C_{max}/k_B$ &
$T_{C,max}/T_{c,sq}$ \\ \hline\hline
2 & $FBC_y$ & 1 & 2.40 & 0.439 & 0.367 \\ \hline
2 & $FBC_y$ & 2 & 1.33 & 0.631 & 0.663 \\ \hline
2 & $FBC_y$ & 3 & 1.17 & 0.755 & 0.753 \\ \hline\hline
2 & $PBC_y$ & 2 & 1.00 & 0.529 & 0.8805 \\ \hline
2 & $PBC_y$ & 3 & 1.88 & 0.366 & 0.469 \\ \hline\hline
3 & $FBC_y$ & 1 & 2.23 & 0.241 & $-$  \\ \hline
3 & $FBC_y$ & 2 & 1.785 & 0.3325& $-$ \\ \hline
3 & $FBC_y$ & 3 & 1.70 & 0.369 & $-$ \\ \hline\hline
3 & $PBC_y$ & 2 & 1.33 & 0.287 & $-$  \\ \hline
3 & $PBC_y$ & 3 & 2.62 & 0.498 & $-$ \\ \hline\hline
4 & $FBC_y$ & 1 & 2.16 & 0.166 & $-$  \\ \hline
4 & $FBC_y$ & 2 & 1.96 & 0.238 & $-$ \\ \hline
4 & $FBC_y$ & 3 & 1.92 & 0.263 & $-$ \\ \hline\hline
4 & $PBC_y$ & 2 & 1.49 & 0.208 & $-$ \\ \hline
4 & $PBC_y$ & 3 & 2.16 & 0.346 & $-$ \\ \hline\hline
\end{tabular}
\end{center}
\label{cafmtable}
\end{table}

In Figs. \ref{cafmsqpxy3strip} and \ref{cafmsqpxpy3strip} we show plots of the
specific heat for the Potts antiferromagnet on the infinite-length limits of
the $L_y=3$ strips with free and periodic transverse boundary conditions.
Using our exact results, we have calculated similar plots for $L_y=2$, but to
save space we do not show them here.  In Table \ref{cafmtable} we list some
comparative information on $K_{C,max}$ and $C_{max}$ from our previous and
current exact calculations.  In contrast to the situation with the ferromagnet,
$|K_{C,max}|$ does not have a uniform behavior with respect to $q$; for
example, for the strips with free transverse boundary conditions, for $L_y=1$,
it decreases with increasing $q$ while for $L_y=2$ and $L_y=3$ it increases
with increasing $q$.  For the strips with periodic transverse boundary
conditions, $|K_{C,max}|$ increases with increasing $q$ for $L_y=2$ but is not
a monotonic function of $q$ for $L_y=3$.  Note that for the $q=2$ Ising value
on the $L_y=3$ strip with periodic boundary conditions the frustration affects
the value of $K_{C,max}$; this would also be true on wider strips of this type
with odd $L_y$. For fixed $q$, as a function of $L_y$, we find that for the
strips with free transverse boundary conditions, $|K_{C,max}|$ decreases and
$C_{max}$ increases as $L_y$ increases; for the strips with periodic
boundary conditions these trends also apply for $q \ge 3$, but for $q=2$, one
encounters the effects of frustration for $L_y=3$ and, more generally, for odd
$L_y$.  Since the $q=3$ Potts antiferromagnet has a zero-temperature critical 
point on the square lattice, one can make use of the relation (\ref{frel}) 
from conformal field theory.  In this case the model exhibits ground 
state entropy without frustration, and, since $\lim_{T \to 0} \beta U = 0$, 
the general relation $f=-\beta U +
S/k_B$ (i.e., $F=U-TS$) reduces to $f = S_0/k_B$. Hence, eq. (\ref{frel}) 
becomes
\beq
S_{strip,L_y} = S_{bulk} + \frac{S_{surf.}}{L_y} + \frac{\Delta}{L_y^2} +
O(L_y^{-3}) \ . 
\label{srel}
\eeq
Using the exact calculation of the ground state entropy for the $L_y=4$ strip
with cylindrical boundary conditions \cite{strip2}, it was noted earlier
\cite{w2d} that this relation is satisfied quite well for the known value of
the central charge, $c=1$. 

\subsubsection{$T \to 0$ Limit}

As regards the low-temperature expansions of the internal energy and specific
heat for the antiferromagnetic Potts model on these strips, these depend on $q$
in a more complicated manner than the analogous expansions for the
ferromagnetic case, as was already clear from eqs. (6.35)-(6.37) and (6.39) in
\cite{a}.  For the Ising value $q=2$ on strips of the square lattice with free
transverse boundary conditions, the limit $L_x \to \infty$ can always be taken
such that the graphs are bipartite, and hence the correction terms in $U$ and
$C$ are obtained from those for the ferromagnet by the replacement $K \to -K$.
However, for other values of $q$, the dependence is, in general, different.
For example, for $L_y=2$, $C$ has an exponential factor $e^{2K}$ if $q=2$ but
$e^{K}$ if $q \ne 2$ \cite{a}.

\section{Locus ${\cal B}$ for the $L_y=3$ Strip with Cyclic or M\"obius 
Boundary Conditions}

\subsection{${\cal B}$ in the $q$ Plane} 

The singular locus ${\cal B}$ in the $q$ plane was given in \cite{wcyl} for 
the special case $a=0$, i.e., the zero-temperature 
Potts antiferromagnet. This locus separates the $q$ plane into seven regions. 
The first of these is $R_1$, which includes the real intervals $q < 0$ and 
$q > q_c$, where
\beq
q_c=2.33654.. \quad {\rm for} \ \ a=0, \ \ G = sq, \ L_y=3, \ FBC_y, \
(T)PBC_x \ .
\label{qccyclly3}
\eeq
The other regions are $R_2$, including the real interval $2 < q < q_c$; $R_3$
including the real interval $0 < q < 2$; two complex-conjugate (c.c.) phases
$R_4,R_4^*$ centered at $q \simeq 2.4 \pm 0.9i$; and two additional quite
narrow, sliver-like c.c. phases $R_5,R_5^*$ centered at $q \simeq 1.93 \pm
1.70i$.  In region $R_1$ the dominant $\lambda$ is, with our current ordering
of terms, $\lambda_{P,s3c,9}$, so that for either cyclic or M\"obius
longitudinal boundary conditions, 
\beq 
W(sq,3 \times \infty, FBC_y,(T)PBC_x,q)=
(\lambda_{P,s3c,9})^{1/3} \quad {\rm for} \ \ q \in R_1 \ .
\label{wsqly3r1}
\eeq
In region $R_2$, the dominant $\lambda$ is $q-4$, so
\beq
|W(sq,3 \times \infty,FBC_y,(T)PBC_x,q)|=|q-4|^{1/3} 
\quad {\rm for} \quad q \in R_2 \ .
\label{wsqly3r2}
\eeq
Recall that for regions other than $R_1$, only the magnitude of $W$ can be
determined unambiguously \cite{w}.
In region $R_3$, the dominant $\lambda$ is the maximal root of the cubic
(\ref{sqcubic}).  In the complex conjugate pairs of regions ($R_4,R_4^*$) and
($R_5,R_5^*$) the dominant $\lambda$'s are the other two roots of the cubic
(\ref{sqcubic}). 
In each of these regions, $|W|=|\lambda_{P,s3c,j,dom.}|^{1/3}$,
where $\lambda_{P,s3c,j,dom.}$ is the dominant root in the respective region. 

\begin{figure}[hbtp]
\centering
\leavevmode
\epsfxsize=3in
\begin{center}
\leavevmode
\epsffile{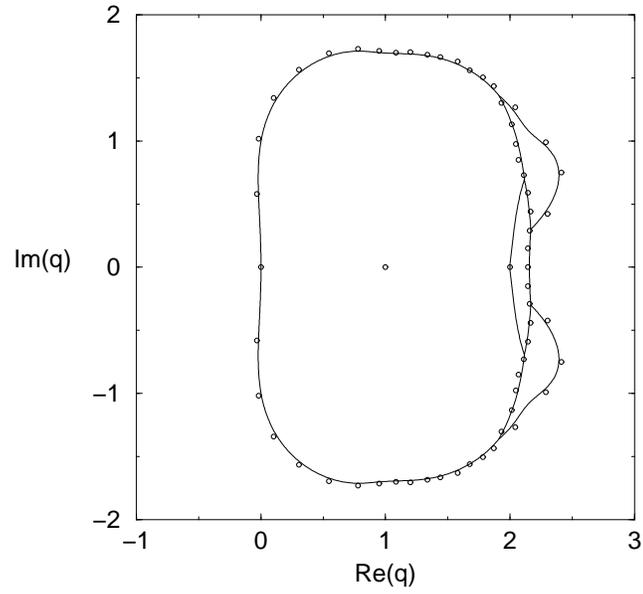}
\end{center}
\caption{\footnotesize{Singular locus ${\cal B}$ in the $q$ plane for the free
energy of the Potts antiferromagnet for the temperature given by $a=0.1$, i.e.,
$K=-\ln 10$, on the width $L_y=3$, infinite-length strip of the square lattice
with cyclic or M\"obius boundary conditions.  Partition function zeros for the
cyclic strip with length $L_x=20$ and thus $n=60$ vertices are shown for
comparison.}}
\label{sqpxy3a0p1}
\end{figure}

\begin{figure}[hbtp]
\centering
\leavevmode
\epsfxsize=3in
\begin{center}
\leavevmode
\epsffile{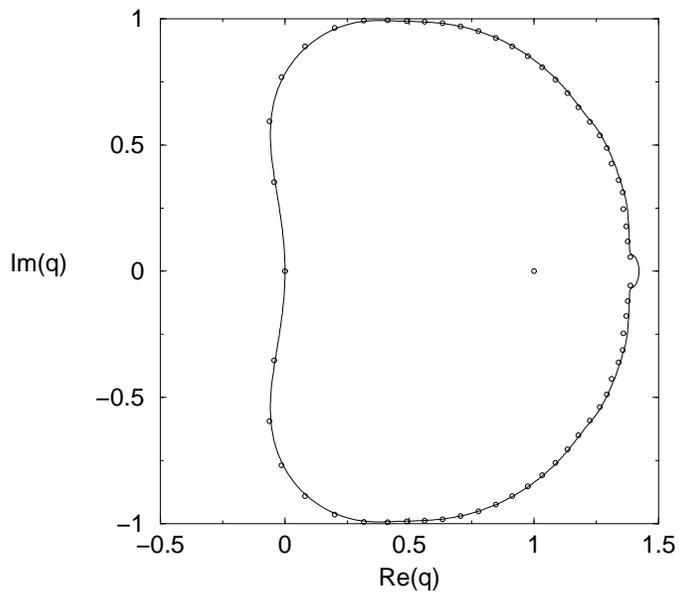}
\end{center}
\caption{\footnotesize{Singular locus ${\cal B}$ in the $q$ plane for the free
energy of the Potts antiferromagnet for the temperature given by $a=0.5$, i.e.,
$K=-\ln 2$, on the width $L_y=3$, infinite-length strip of the square lattice
with cyclic or M\"obius boundary conditions.  Partition function zeros for the
cyclic strip with length $L_x=20$ and thus $n=60$ vertices are shown for
comparison.}}
\label{sqpxy3a0p5}
\end{figure}

In Figs. \ref{sqpxy3a0p1} and \ref{sqpxy3a0p5} we show plots of the locus
${\cal B}_q$ for the Potts antiferromagnet at the values of temperature given
by $a=0.1$, i.e., $K=-\ln 10$, and $a=0.5$, i.e., $K=-\ln 2$.  For comparison,
typical sets of partition function zeros in the $q$ plane are shown for long
finite strips.  As $a$ increases from 0 to 1, i.e., the temperature increases
from 0 to infinity for the Potts antiferromagnet, the boundary ${\cal B}_q$
contracts and finally shrinks to a point at the origin in the $q$ plane.  In
the interval $0 \le a \le a_{crc}$, where $a_{crc}=0.181720..$, the rightmost
part of the boundary and hence also $q_c$ sweep to the left, while the crossing
at $q=2$ remains; at $a=a_{crc}$, the rightmost part of the boundary coincides
with $q=2$, i.e., $q_c=2$.  As $a$ increases further in the interval $a_{crc}
\le a \le 1$, there are only two regions, $R_1$ and $R_3$ that include real
intervals.  For this full range of temperatures in the Potts antiferromagnet,
$0 \le a \le 1$, the boundary ${\cal B}$ always passes through $q=0$.

\begin{figure}[hbtp]
\centering
\leavevmode
\epsfxsize=3in
\begin{center}
\leavevmode
\epsffile{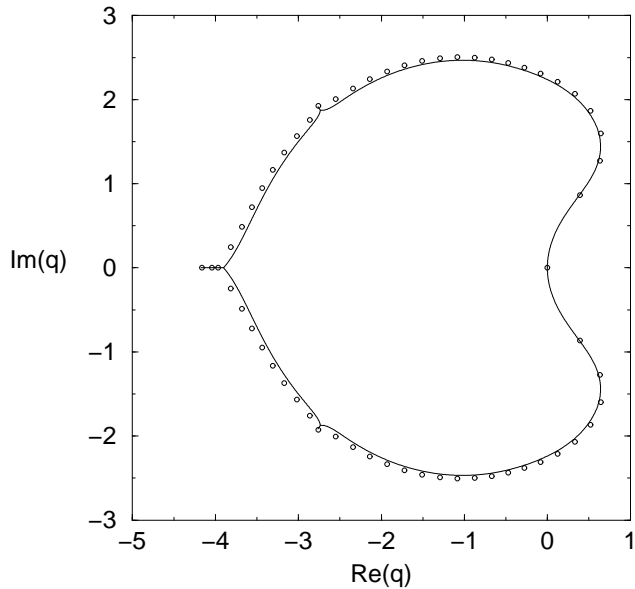}
\end{center}
\caption{\footnotesize{Singular locus ${\cal B}$ in the $q$ plane for the free
energy of the Potts ferromagnet for the temperature given by $a=2$, i.e.,
$K=\ln 2$, on the width $L_y=3$, infinite-length strip of the square lattice
with cyclic or M\"obius boundary conditions.  Partition function zeros for the
cyclic strip with length $L_x=20$ and thus $n=60$ vertices are shown for
comparison.}}
\label{sqpxy3a2}
\end{figure}

In Fig. \ref{sqpxy3a2} we show the locus ${\cal B}_q$ for a typical
ferromagnetic value, $a=2$, i.e. $K = \ln 2$.  One can discern very small
prongs extending in northwest and southwest directions from the main curve at
$q \simeq -2.74 \pm 1.88i$, and a short line segment on the negative real $q$ 
axis. Two general features are that the locus ${\cal B}$ (i) passes through
$q=0$ and (ii) does not cross the positive real.

\subsection{${\cal B}$ in the $u$ Plane}

\begin{figure}[hbtp]
\centering
\leavevmode
\epsfxsize=3in
\begin{center}
\leavevmode
\epsffile{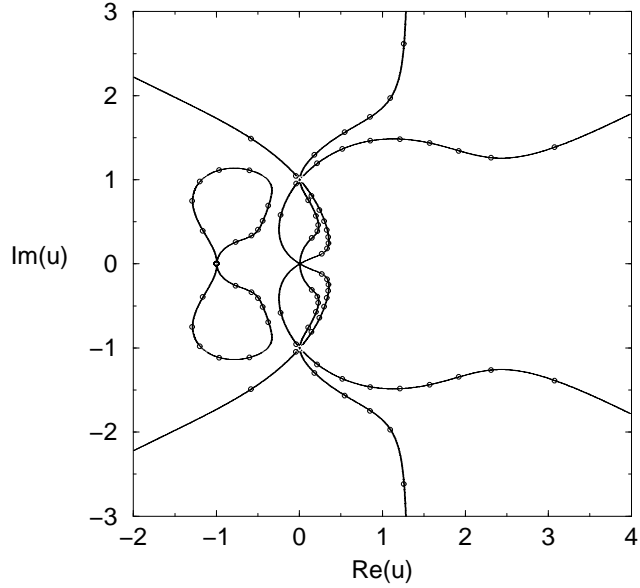}
\end{center}
\caption{\footnotesize{Singular locus ${\cal B}$ in the $u$ plane for the free
energy of the Potts model for $q=2$, on the width $L_y=3$, infinite-length
strip of the square lattice with cyclic or M\"obius boundary conditions.
Partition function zeros for the cyclic strip with length $L_x=20$ and hence
$e=100$ are shown for comparison.}}
\label{sqpxy3q2}
\end{figure}

\begin{figure}[hbtp]
\centering
\leavevmode
\epsfxsize=3in
\begin{center}
\leavevmode
\epsffile{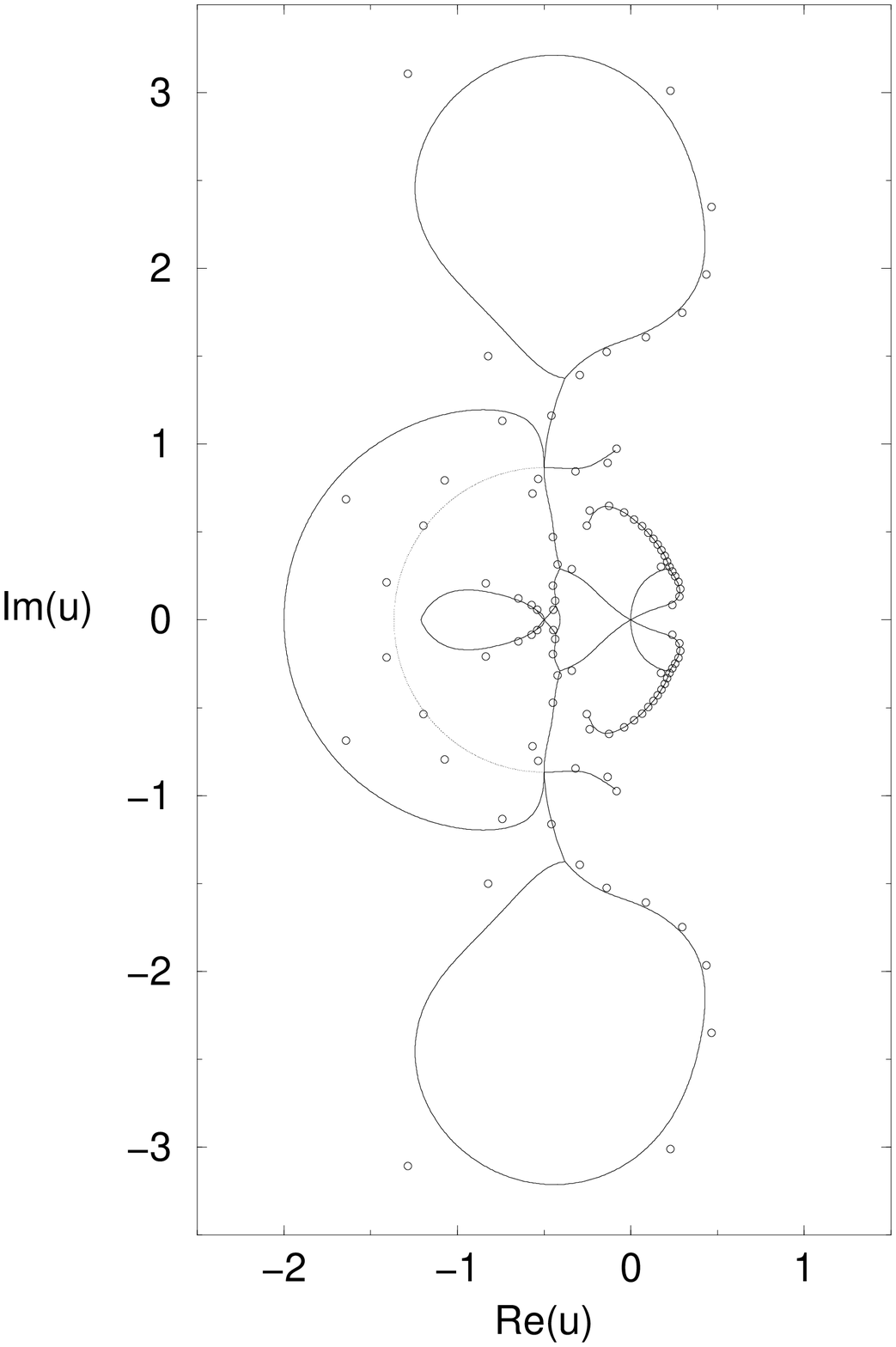}
\end{center}
\caption{\footnotesize{Singular locus ${\cal B}$ in the $u$ plane for the free
energy of the Potts model for $q=3$, on the width $L_y=3$, infinite-length
strip of the square lattice with cyclic or M\"obius boundary conditions.
Partition function zeros for the cyclic strip with length $L_x=20$ and hence
$e=100$ are shown for comparison.}}
\label{sqpxy3q3}
\end{figure}

\begin{figure}[hbtp]
\centering
\leavevmode
\epsfxsize=3in
\begin{center}
\leavevmode
\epsffile{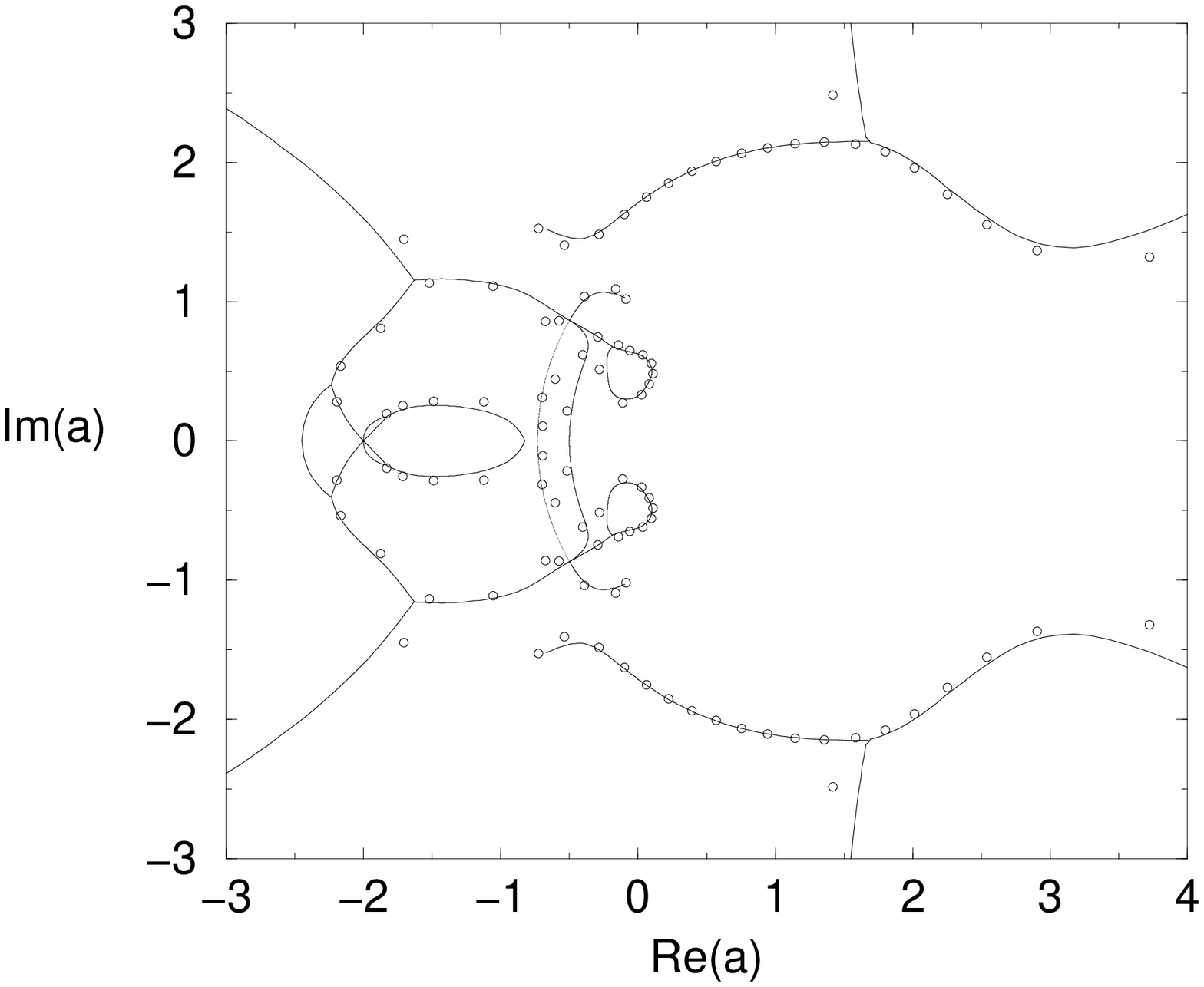}
\end{center}
\caption{\footnotesize{Singular locus ${\cal B}$ in the $a=u^{-1}$ plane for
the free energy of the Potts model for $q=3$, on the width $L_y=3$,
infinite-length strip of the square lattice with cyclic or M\"obius boundary
conditions.  Partition function zeros for the cyclic strip with length $L_x=20$
and hence $e=100$ are shown for comparison.}}
\label{sqpxy3aq3}
\end{figure}

\begin{figure}[hbtp]
\centering
\leavevmode
\epsfxsize=3in
\begin{center}
\leavevmode
\epsffile{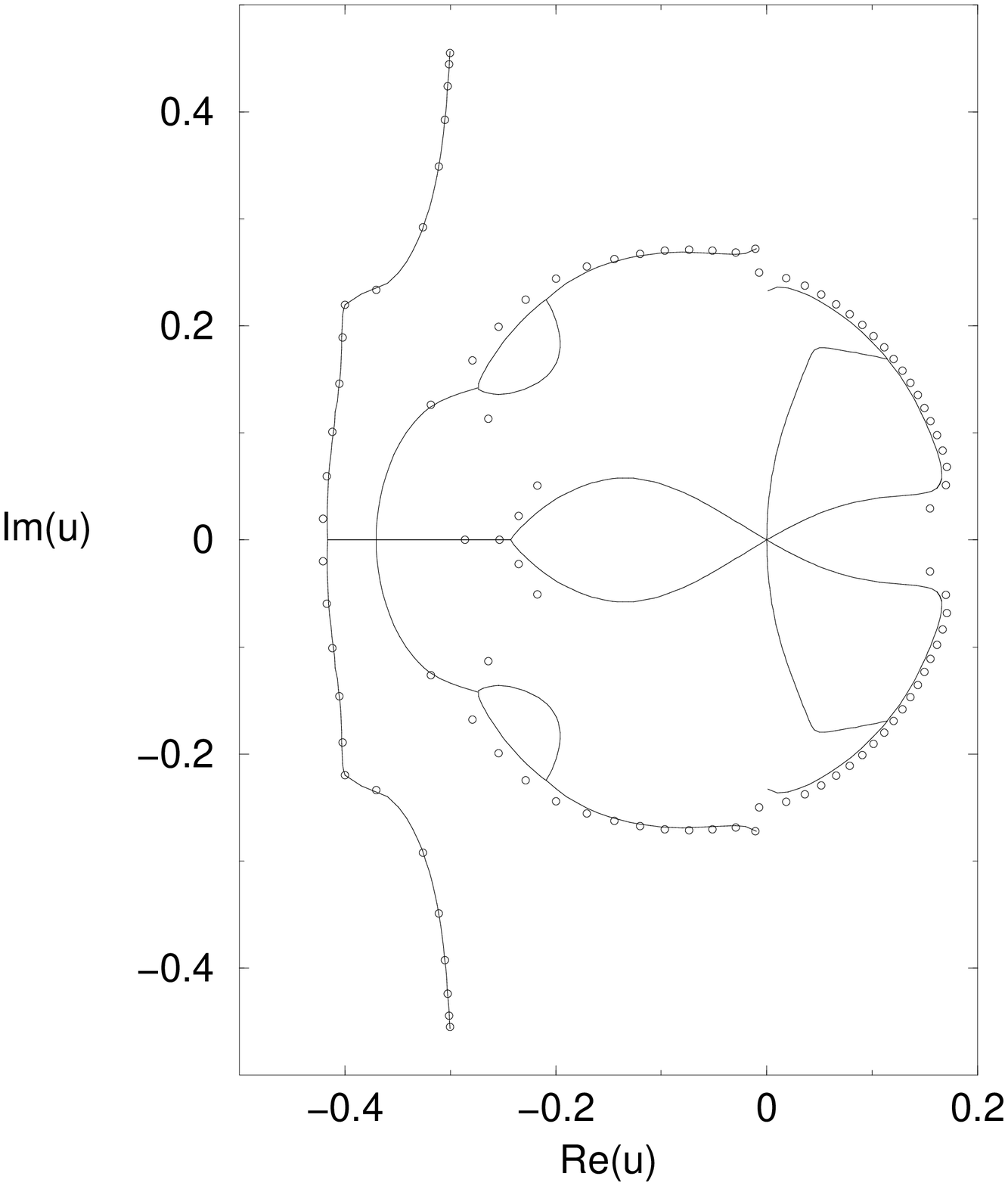}
\end{center}
\caption{\footnotesize{Singular locus ${\cal B}$ in the $u$ plane for the free
energy of the Potts model for $q=10$, on the width $L_y=3$, infinite-length
strip of the square lattice with cyclic or M\"obius boundary conditions.
Partition function zeros for the cyclic strip with length $L_x=20$ and hence
$e=100$ are shown for comparison.}}
\label{sqpxy3q10}
\end{figure}

In Figs. \ref{sqpxy3q2}-\ref{sqpxy3q10} we show plots of the singular locus 
${\cal B}$ for the Potts model free energy for $q=2,3$, and 10, respectively, 
in the infinite-length limit of the $L_y=3$ strip graph with cyclic or 
M\"obius boundary conditions.  Zeros of the partition function 
are shown for comparison.  There are several interesting features of these 
plots.  In each, there are six curves forming three branches on ${\cal B}$ 
that intersect at the origin, $u=0$, at the angles $\pm \pi/6$, $\pm \pi/2$,
and $\pm 5\pi/6$.  In general, the density of complex-temperature Fisher 
zeros along the curves comprising ${\cal B}$ in the
vicinity of a generic singular point $u_s$ behaves as \cite{fisher,abe}
\beq
g \sim |u-u_s|^{1-\alpha_s}
\label{gdensity}
\eeq
where $\alpha_s$ ($\alpha_s'$) denotes the
corresponding specific heat exponent for the approach to $u_s$ from within the
CTE PM (FM) phase.  Thus, for a continuous, second-order transition, with
$\alpha_s < 1$, this density vanishes as one approaches the critical point
$u_s$ along ${\cal B}$. The specific heat for these infinite-length,
finite-width strips has an essential zero at $T=0$ which, if expressed in 
terms of an algebraic specific exponent $\alpha$,
corresponds to $\alpha=-\infty$ at $u=u_s=0$.  Substituting this into
(\ref{gdensity}), it follows that the density vanishes rapidly as one
approaches the origin $u=0$ along the curves forming ${\cal B}_u$.  This is
evident in Figs. \ref{sqpxy3q2}-\ref{sqpxy3q10}.  As we have discussed in our 
earlier 
works on ${\cal B}$ in the $q$ and $u$ plane, any mapping of the regions in
these respective planes is done with finite resolution and does not exclude 
the possibility of extremely small regions.  That such regions can occur was
evident, e.g., in \cite{wcyl}.

In earlier work \cite{is1d,a,ta,ka,hca}, it was shown that although the
physical thermodynamic properties of a discrete spin model are, in general,
different in quasi-1D systems such as infinite-width, finite-width strips, and
in higher dimensions, nevertheless, exact solutions for ${\cal B}_u$ in 
quasi-1D systems can give insight into ${\cal B}_u$ in 2D.  This was shown, in
particular, for the $q=2$ Ising special case of the Potts model, where the
comparison can be made rigorously since the model is exactly solvable in 2D.
Our current results extend this comparison to greater width. 
For example, in the complex-temperature phase diagram
of the Ising model on the square lattice, the singular locus ${\cal B}$ forms
two circles $u = \pm 1 + \sqrt{2}e^{i\phi}$, $0 \le \phi < 2\pi$
\cite{fisher,chisq}.  This locus thus exhibits complex-conjugate multiple
points at $u=\pm i$, where four curves forming two branches on ${\cal B}$
intersect.  This was also true of the Ising model on the infinite-length, width
$L_y=2$ cyclic or M\"obius strips of the square lattice \cite{bcc,a}. 

For the $q=2$ (Ising) case shown in Fig. \ref{sqpxy3q2}, the locus ${\cal B}$
is invariant under the inversion mapping $u \to 1/u = a$.  This is a
consequence of the fact that the infinite-length limit can be taken on a
sequence of bipartite graphs.  It follows that the six curves extending outward
to infinity in the $u$ plane cross at the origin of the $a$ plane.  Thus, the
points $u=0$ and $a=0$ are multiple points on ${\cal B}$.  There are also
complex-conjugate multiple points at $u=\pm i$, where again six curves forming
three branches of ${\cal B}$ cross each other.  Finally, there is another
multiple point at $u=-1$ where four curves forming two branches intersect.
Although the locus ${\cal B}$ does not appear to contain any arc endpoints for
$q=2$, such endpoints are present for the other illustrative values of $q$,
namely $q=3$ and $q=10$.  

In the $q=3$ case, the intersection points of the various curves include the
complex-conjugate pair $u=e^{\pm 2\pi i/3}$ as well as a negative real point,
$u=-1/2$. The intersection points include both those where several curves cross
and those where curves come together in a $T$-type intersection point.  A
previous analytic discussion of such $T$-type intersection points in
complex-temperature phase diagrams was given in \cite{z6}.  Besides the origin,
$u=0$, the locus ${\cal B}$ crosses the real $u$ axis also at the three points
$u=-2$, $u \simeq -1.37$, and $u \simeq -1.21$.  We recall that for $q=3$ Potts
model on the infinite-length, width $L_y=2$ cyclic or M\"obius strip of the
square lattice, ${\cal B}$ also included multiple points at $u=e^{\pm 2\pi
i/3}$ and crossed the real axis at $u=-2$ and $u=-1/2$ (as well as the origin)
\cite{a}.  These results for the $L_y=2$ and $L_y=3$ cyclic/M\"obius strips
suggest that it is likely that for the infinite 2D square lattice, ${\cal B}$
passes through the points $u=e^{\pm 2 \pi i/3}$.  Calculations of Fisher zeros
on finite lattices show considerable scatter in the region $Re(a) < 0$
\cite{mbook,wuz,pfef} but are consistent with this suggestion.  The crossings
at $u=-1/2$, i.e., $a=-2$, for ${\cal B}$ on these $L_y=2,3$ strips are of
interest since, given that the $q=3$ Potts antiferromagnet has a
zero-temperature critical point on the square lattice, so that the singular
locus ${\cal B}$ passes through $a=0$, it follows by duality that this locus
also passes through $a=-2$ \cite{pfef}.  Since some features of ${\cal B}$ are
clearer when it is shown in the $a=u^{-1}$ plane, we do this in
Fig. \ref{sqpxy3aq3}.

The plot of ${\cal B}$ for $q=10$ plot in Fig. \ref{sqpxy3q10} shows another
interesting feature: a portion of this locus, and the associated partition
function zeros exhibit an approximately circular shape.  Of course the actual
locus is considerably more complicated than a circle.  However, in this
context, one may recall that in the $q \to \infty$ limit, the
complex-temperature phase boundary is $|\zeta|=1$, where 
$\zeta=(a-1)/\sqrt{q}$ \cite{wuetal}, which, for large $q$, yields the locus 
$|u| \to 1/\sqrt{q}$ in the $u$ plane.  Thus, aside from the features of 
${\cal B}$ that reflect the quasi-1D nature of the strip, such as the curves
passing through $u=0$, the width $L_y=3$ is large enough so that one can begin
to see this feature of the $q \to \infty$ limit.

\section{Locus ${\cal B}$ for $L_y=2$ Strip with Torus/Klein Bottle Boundary
Conditions} 

We find that the locus ${\cal B}_q$ for the (infinite-length limit of the)
$L_y=2$ strip with torus or Klein bottle boundary conditions is generally
similar to that for the same strip with cyclic or M\"obius boundary conditions
\cite{a}.  For the finite-temperature ferromagnet this is again true.  In
particular, ${\cal B}_q$ for the ferromagnet does not cross the positive real
$q$ axis.

\begin{figure}[hbtp]
\centering
\leavevmode
\epsfxsize=3in
\begin{center}
\leavevmode
\epsffile{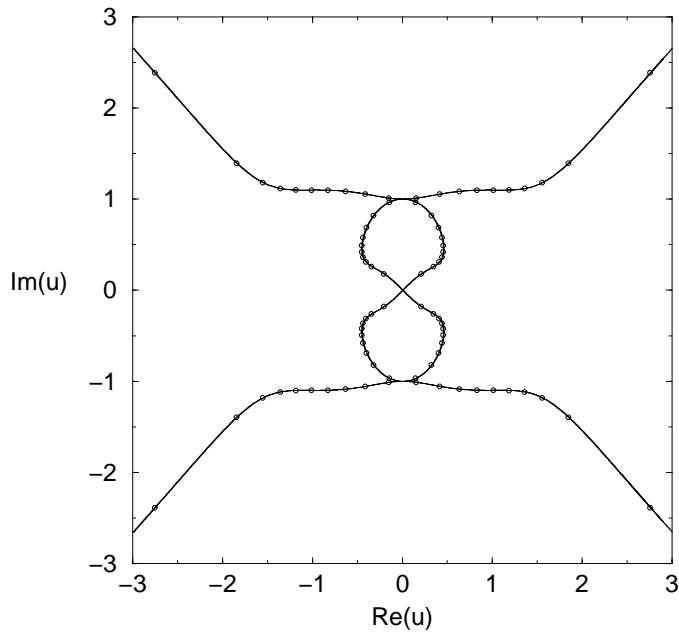}
\end{center}
\caption{\footnotesize{Singular locus ${\cal B}$ in the $u$ plane for the free
energy of the Potts model for $q=2$, on the width $L_y=2$, infinite-length
strip of the square lattice with torus or Klein bottle boundary conditions.
Partition function zeros for the toroidal strip with length $L_x=20$ and hence
$e=80$ are shown for comparison.}}
\label{sqpxpy2q2}
\end{figure}

\begin{figure}[hbtp]
\centering
\leavevmode
\epsfxsize=3in
\begin{center}
\leavevmode
\epsffile{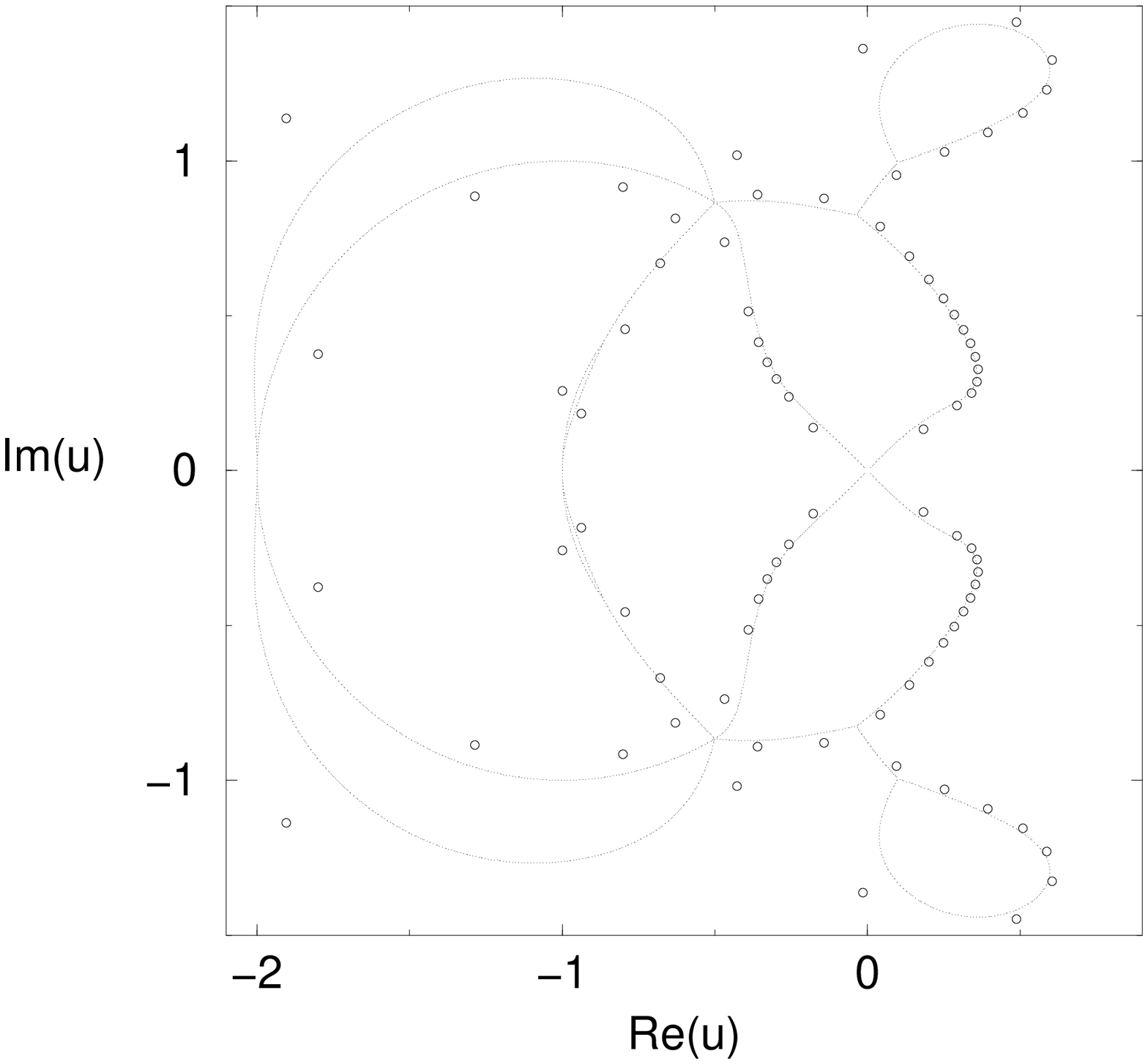}
\end{center}
\caption{\footnotesize{Singular locus ${\cal B}$ in the $u$ plane for the free
energy of the Potts model for $q=3$, on the width $L_y=2$,
infinite-length strip of the square lattice with torus or Klein bottle boundary
conditions. Partition function zeros for the toroidal strip with length
$L_x=20$ ($e=80$) are shown for comparison.}}
\label{sqpxpy2q3}
\end{figure}

In Figs. \ref{sqpxpy2q2} and \ref{sqpxpy2q3} we plot ${\cal B}_u$ for $q=2$ and
$q=3$.  These loci contain four curves forming two branches that cross at the
origin, with angles $\pm \pi/4$ and $\pm 3\pi/4$.  Comparing the locus for
$q=2$ plot with the corresponding locus in the case of the infinite-length
$L_y=2$ cyclic or M\"obius strip of the square lattice \cite{a,ta} we observe
that here the locus is symmetric under the map $u \to -u$, whereas for the
cyclic or M\"obius strip it is not.  The reason for this is that the torus or
Klein bottle strip has an even coordination number, $\Delta=4$, whereas the
cyclic or M\"obius strip has an odd coordination number, $\Delta=3$ (both are
$\Delta$-regular graphs).  The $q=3$ locus contains intersection points at $u =
e^{ \pm 2 \pi i/3}$ as well as two pairs of $T$ (equivalently, $Y$) type
intersection points, and crosses the negative real axis at $u=-1$ and $u=-2$.

\section{Locus ${\cal B}$ for $L_y=3$ Strip with Torus/Klein Bottle Boundary
Conditions}

\subsection{ ${\cal B}$ in the $q$ Plane}

\begin{figure}[hbtp]
\centering
\leavevmode
\epsfxsize=3in
\begin{center}
\leavevmode
\epsffile{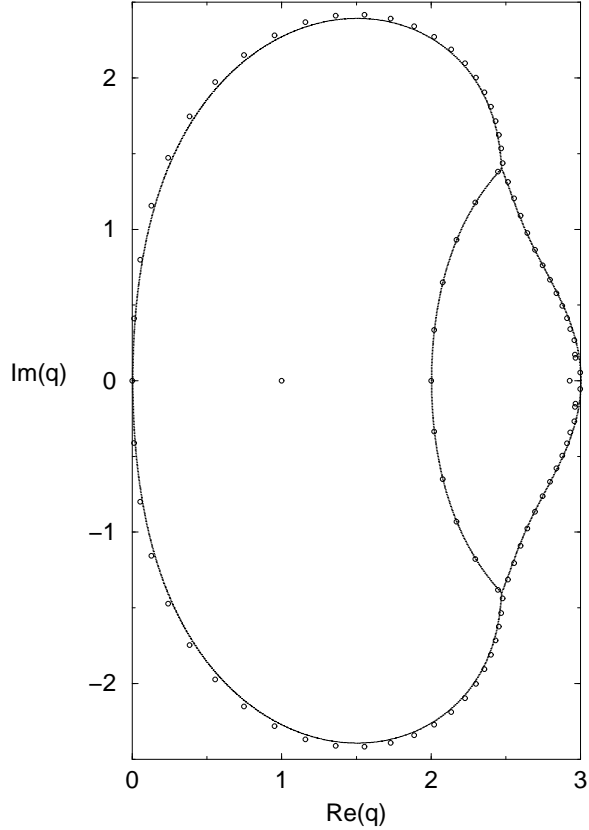}
\end{center}
\caption{\footnotesize{Singular locus ${\cal B}$ in the $q$ plane for the free
energy of the zero-temperature Potts antiferromagnet on the $L_y=3$ strip of
the square lattice with torus or Klein bottle boundary conditions.  Chromatic
zeros for the toroidal strip with length $L_x=30$ and thus $n=90$ vertices are
shown for comparison.}}
\label{sqpxpy3}
\end{figure}

\begin{figure}[hbtp]
\centering
\leavevmode
\epsfxsize=3in
\begin{center}
\leavevmode
\epsffile{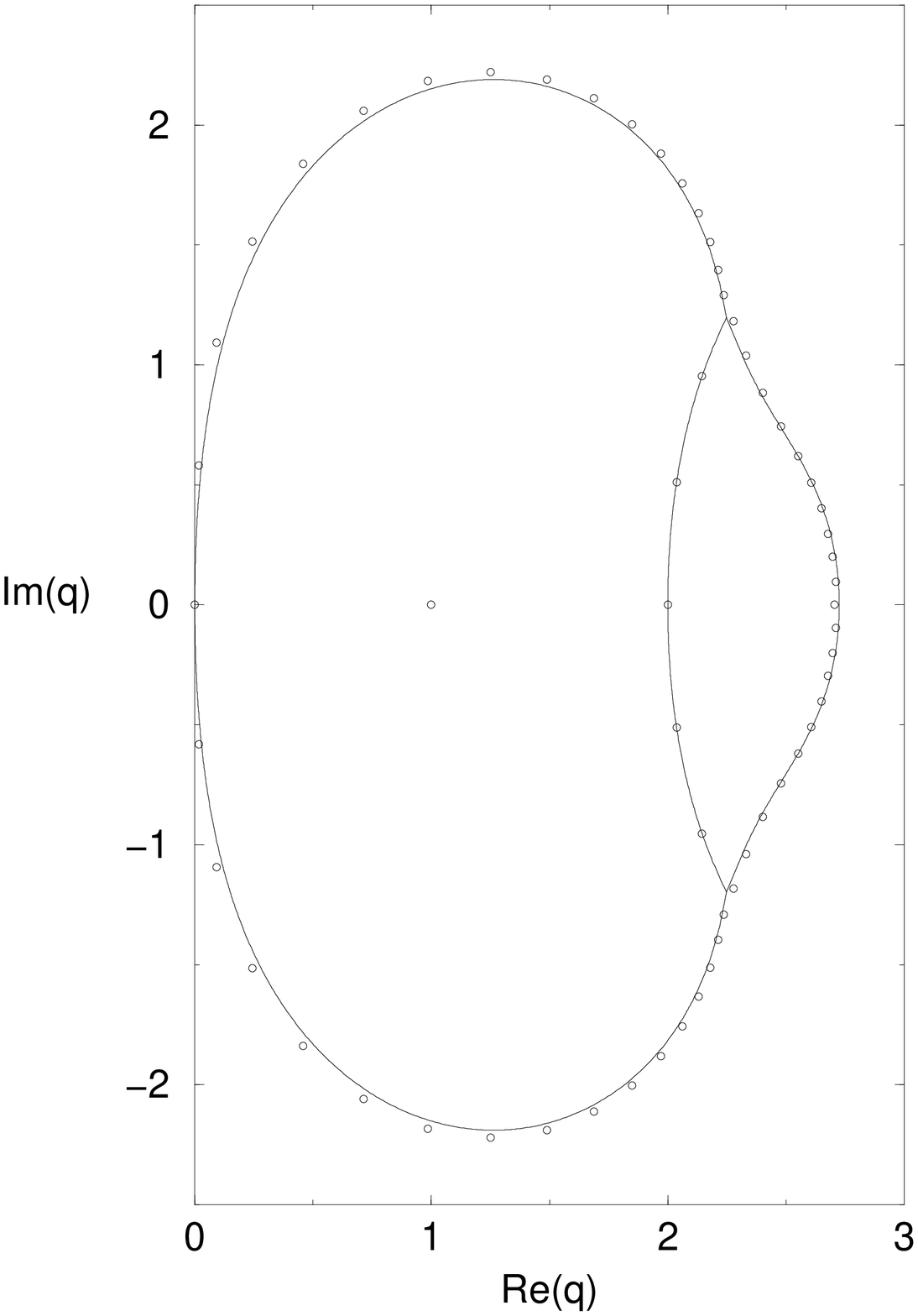}
\end{center}
\caption{\footnotesize{Singular locus ${\cal B}$ in the $q$ plane for the free
energy of the Potts antiferromagnet for the temperature given by $a=0.1$, i.e.,
$K=-\ln 10$, on the width $L_y=3$, infinite-length strip of the square lattice
with torus or Klein bottle boundary conditions.  Partition function zeros for
the toroidal strip with length $L_x=20$ and thus $n=60$ vertices are shown for
comparison.}}
\label{sqpxpy3a0p1}
\end{figure}

\begin{figure}[hbtp]
\centering
\leavevmode
\epsfxsize=3in
\begin{center}
\leavevmode
\epsffile{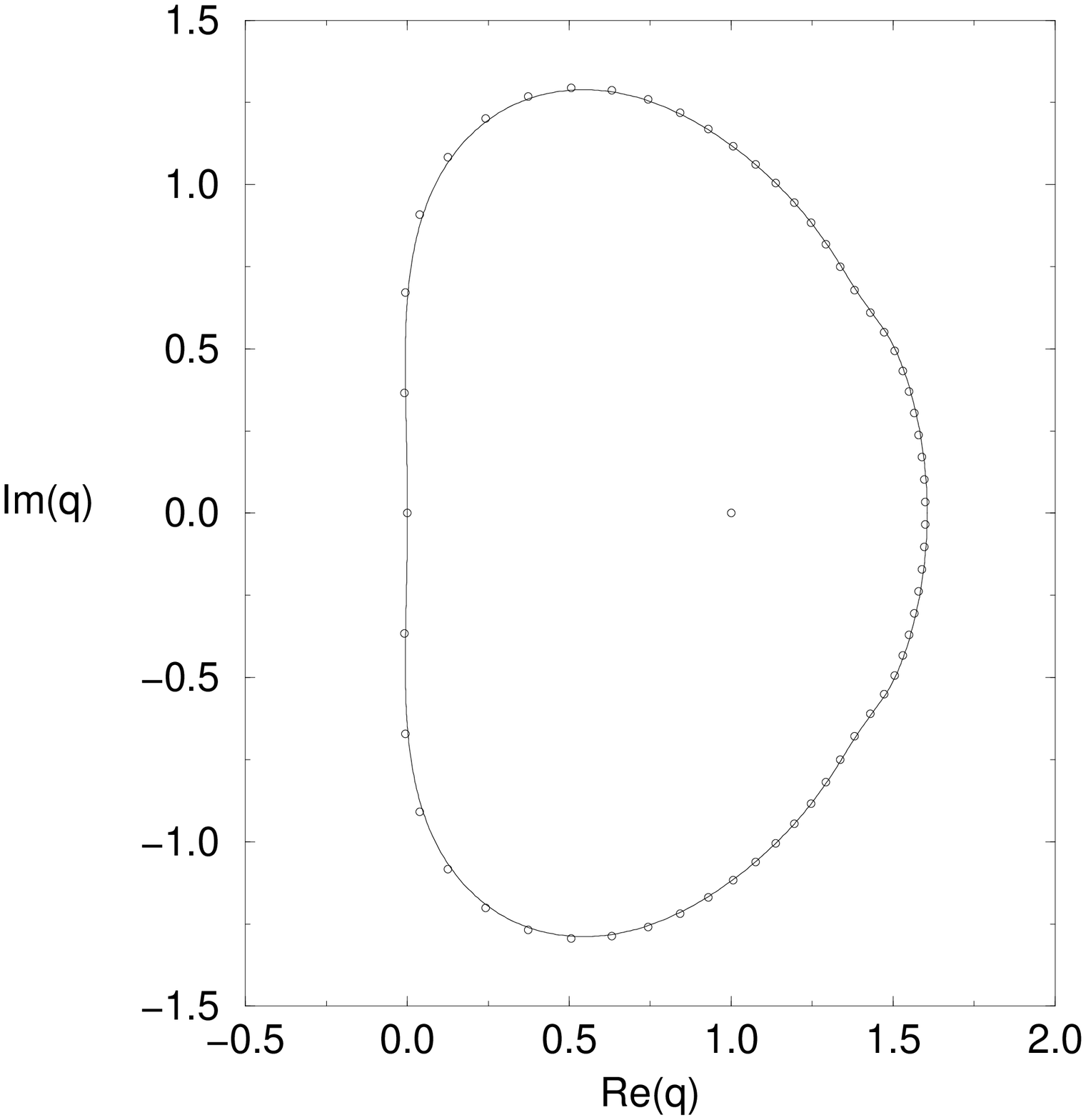}
\end{center}
\caption{\footnotesize{Singular locus ${\cal B}$ in the $q$ plane for the free
energy of the Potts antiferromagnet for the temperature given by $a=0.5$, i.e.,
$K=-\ln 2$, on the width $L_y=3$, infinite-length strip of the square lattice
with torus or Klein bottle boundary conditions.  Partition function zeros for
the toroidal strip with length $L_x=20$ and thus $n=60$ vertices are shown for
comparison.}}
\label{sqpxpy3a0p5}
\end{figure}

For comparison with our finite-temperature results, we show in
Fig. \ref{sqpxpy3} the locus ${\cal B}$ for the infinite-length limit of the
$L_y=3$ strip with torus or Klein bottle boundary conditions in the $q$ plane
for $a=0$, i.e., the zero-temperature Potts antiferromagnet, on the $L_y=3$
strip of the square lattice with torus or Klein bottle boundary conditions
\cite{tk}.  This locus separates the $q$ plane into three regions: $R_1$
including the real intervals $q > q_c$, where $q_c=3$, and $q < 0$ and
extending to arbitrarily large $|q|$; $R_2$, including the real interval $2 < q
< 3$, and (iii) $R_3$, including the real interval $0 < q < 2$.  Thus, ${\cal
B}_q$ crosses the real $q$ axis at $q=0,2$, and 3.
For the same model at the finite values of temperature given by 
$a=0.1$ and $a=0.5$, we show ${\cal B}_q$ in Figs. \ref{sqpxpy3a0p1} and 
\ref{sqpxpy3a0p5}. 
As $a$ increases from 0 to 1, i.e., the temperature increases from 0 to
infinity for the Potts antiferromagnet, the boundary ${\cal B}_q$ contracts to
a point at the origin in the $q$ plane.  In the interval $0 \le a \le a_{cr}$,
where $a_{cr}=0.3488338...$, which is a root of the equation
\beqs & &
2z^{12}+13z^{11}+21z^{10}+38z^9+46z^8-11z^7+40z^6 \cr\cr 
& & -17z^5+44z^4-70z^3+35z^2-17z+4=0 \ .
\label{eqacr}
\eeqs
The boundary between regions $R_2$ and $R_3$ remains fixed at $q=2$ and
the 
right-most part of the boundary, passing through $q_c$ and separating $R_1$ and
$R_2$ moves leftward, and crosses $q=2$ at $a=a_{cr}$.  In the interval $a_{cr}
< a < 1$, there are only two regions, $R_1$ and $R_3$, that contain intervals 
of the real axis, and $q_c$ continues to decrease from 2 to 0. 

\begin{figure}[hbtp]
\centering
\leavevmode
\epsfxsize=3in
\begin{center}
\leavevmode
\epsffile{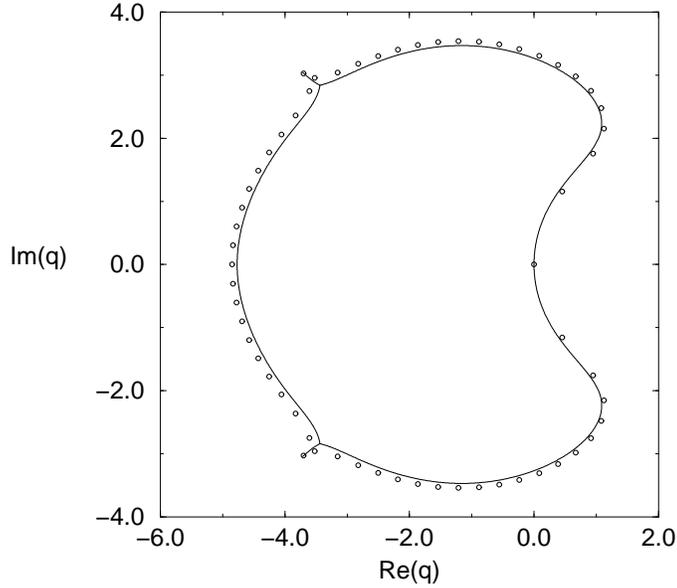}
\end{center}
\caption{\footnotesize{Singular locus ${\cal B}$ in the $q$ plane for the free
energy of the Potts ferromagnet for the temperature given by $a=2$, i.e.,
$K=\ln 2$, on the width $L_y=3$, infinite-length strip of the square lattice
with torus or Klein bottle boundary conditions.  Partition function zeros for
the toroidal strip with length $L_x=20$ and thus $n=60$ vertices are shown for
comparison.}}
\label{sqpxpy3a2}
\end{figure}

In Fig. \ref{sqpxpy3a2} we show the locus ${\cal B}_q$ for a typical
ferromagnetic value, $a=v+1=2$, i.e. $K = \ln 2$.  As is evident in this
figure, ${\cal B}$ includes a complex-conjugate pair of prongs extending
outward from the main curve.  

\subsection{ ${\cal B}$ in the $u$ Plane} 

\begin{figure}[hbtp]
\centering
\leavevmode
\epsfxsize=3in
\begin{center}
\leavevmode
\epsffile{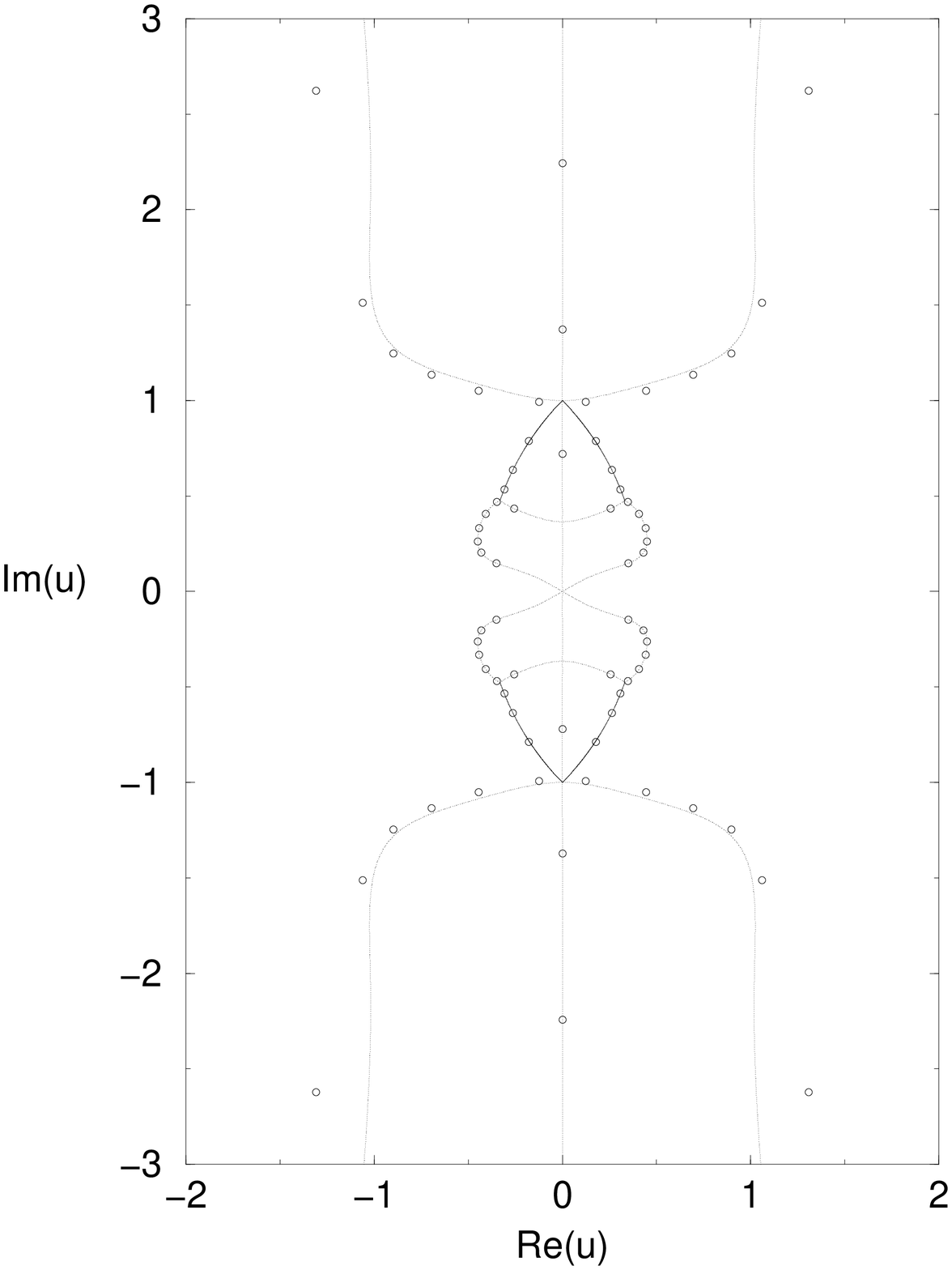}
\end{center}
\caption{\footnotesize{Singular locus ${\cal B}$ in the $u$ plane for the free
energy of the Potts model for $q=2$, on the width $L_y=3$,
infinite-length strip of the square lattice with torus or Klein bottle boundary
conditions.  Partition function zeros for the toroidal strip with length
$L_x=15$ and hence $e=90$ are shown for comparison.}}
\label{sqpxpy3q2}
\end{figure}

\begin{figure}[hbtp]
\centering
\leavevmode
\epsfxsize=3in
\begin{center}
\leavevmode
\epsffile{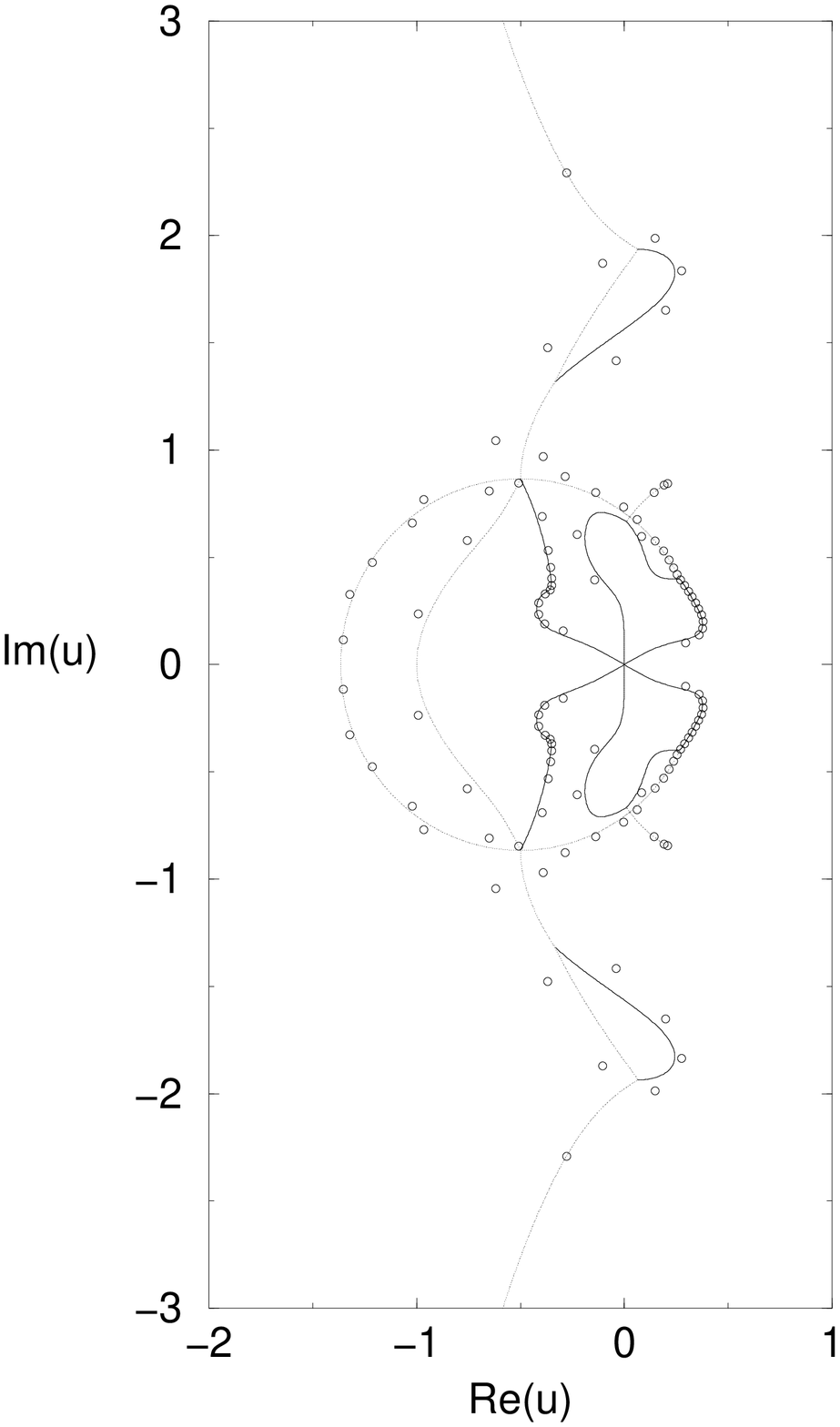}
\end{center}
\caption{\footnotesize{Singular locus ${\cal B}$ in the $u$ plane for the free
energy of the Potts model for $q=3$, on the width $L_y=3$,
infinite-length strip of the square lattice with torus or Klein bottle boundary
conditions.  Partition function zeros for the toroidal strip with length
$L_x=20$ and hence $e=120$ are shown for comparison.}}
\label{sqpxpy3q3}
\end{figure}

\begin{figure}[hbtp]
\centering
\leavevmode
\epsfxsize=3in
\begin{center}
\leavevmode
\epsffile{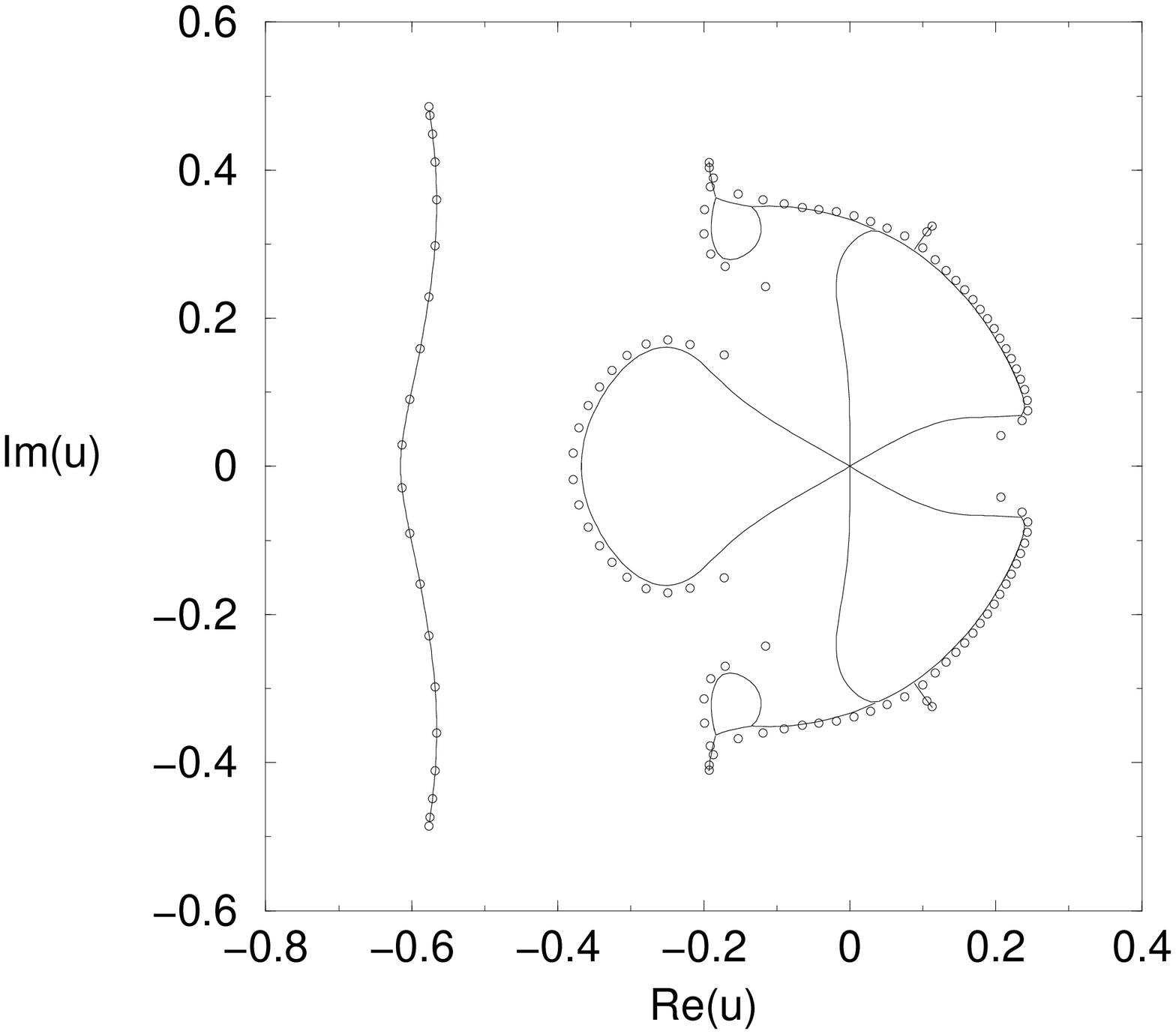}
\end{center}
\caption{\footnotesize{Singular locus ${\cal B}$ in the $u$ plane for the free
energy of the Potts model for $q=10$, on the width $L_y=3$,
infinite-length strip of the square lattice with torus or Klein bottle boundary
conditions.  Partition function zeros for the toroidal strip with length
$L_x=20$ and hence $e=120$ are shown for comparison.}}
\label{sqpxpy3q10}
\end{figure}

We show here plots of the singular locus ${\cal B}$ and partition function 
zeros in the $u=1/a$ plane for several illustrative values of $q$, namely, 
$q=2$, $q=3$, and $q=10$ in Figs. \ref{sqpxpy3q2}- \ref{sqpxpy3q10}.

Some of the features of these plots are similar to those discussed above for
the infinite-length $L_y=3$ strips of the cyclic or M\"obius strips, such as
the sixfold crossing of curves at $u=0$ with density of Fisher zeros rapidly
vanishing near this point, and the additional multiple points at $u=\pm i$ for
$q=2$ and at $u=e^{\pm 2\pi i/3}$ for $q=3$.  Note that in the present case,
for $q=3$, ${\cal B}$ crosses the real $u$ axis at the points $u=-1$ and
$u=-(1+\sqrt{3} / )2 \simeq -1.366...$, in addition to the origin.  In contrast
to the $L_y=3$ cyclic or M\"obius strips, where ${\cal B}$ is compact in the
$u$ plane, here, for the torus or Klein bottle strips, two complex-conjugate
curves on ${\cal B}$ extend to complex infinity in the $u$ plane, i.e., pass
through the origin of the $a$ plane.  This is equivalent to the property shown
in \cite{tk} that ${\cal B}$ for the zero-temperature (i.e., $a=0$) Potts
antiferromagnet passes through $q=q_c=3$.  As was noted in \cite{tk}, this is
interesting, since it indicates that the $q=3$ Potts antiferromagnet has 
a zero-temperature critical point on the infinite-length, width $L_y=3$ 
strips of the square lattice with torus or Klein bottle boundary conditions. 
As we have discussed before, a convenient feature of strips with periodic
longitudinal boundary conditions is that for the $q$-state Potts 
antiferromagnet with a value of $q=q_c$ where the model has a zero-temperature
critical point, this is signalled explicitly by the properties that 
${\cal B}_a$ passes through the origin of the $a$ plane and ${\cal B}_q$ 
passes through $q=q_c$ for $a=0$.  In contrast, if one uses strips with
free longitudinal boundary conditions, this connection is not evident in
general.  Thus, if one uses a cylindrical rather than toroidal or Klein
bottle strip with $L_y=3$, then the boundary ${\cal B}_q$ does not pass through
$q=3$ \cite{strip2}. 

\section{Conclusions} 

In this paper we have presented and discussed exact calculations of the
partition function of the $q$-state Potts model for general $q$ and temperature
on strips of the square lattice of width $L_y=3$ vertices and arbitrary length
$L_x$ with longitudinal boundary conditions of the cyclic, M\"obius, torus, and
Klein bottle types.  In the infinite-length limit we analyzed the resultant
thermodynamic properties and also derived a number of low-temperature
expansions, including results valid for arbitrarily wide strips with free or
periodic transverse boundary conditions.  A number of interesting results were
given for the continuous locus ${\cal B}$ where the free energy is singular in
both the $q$ plane and temperature variable plane.  We also discussed the
related Tutte polynomials.

\vspace{10mm}

Acknowledgment: This research was supported in part by the U. S. NSF
grant PHY-97-22101.

\section{Appendix}

\subsection{Equations for $\lambda_{Z,s3,j}$}

The terms $\lambda_{Z,s3c,j}$ for $10 \le j \le 15$ in eq. (\ref{zgsums3cyc})
and (\ref{zgsums3mb}) are roots of the sixth degree equation
\beq
\xi^6+f_{61}\xi^5+f_{62}\xi^4+f_{63}\xi^3+f_{64}\xi^2+f_{65}\xi+f_{66}=0
\label{eqsix}
\eeq
where
\beq
f_{61}=-v(7v^3+18v^2+v^2q+10qv+2q^2+v^4)
\label{f61}
\eeq
\beqs
& & f_{62}=v^2(q^4+12q^3v+2q^3v^2+v^6q+14v^5q+2v^4q^2+2v^7+19v^6 \cr\cr
& & +78v^4+64v^5+109qv^3+54q^2v^2+66v^4q+20v^3q^2)
\label{f62}
\eeqs
\beqs
& & f_{63}=-v^4(v+q)(2q^4+21q^3v+13q^3v^2+8v^6q+62v^5q+19v^4q^2+16v^7 \cr\cr
& & +79v^6+126v^4+v^8+q^4v+q^3v^3+2v^5q^2+168v^5+168qv^3+91q^2v^2 \cr\cr
& & +182v^4q+79v^3q^2)
\label{f63}
\eeqs
\beqs
& & f_{64}=v^6(v+1)(v+q)^2(q^4+13q^3v+5q^3v^2+2v^6q+26v^5q+7v^4q^2+3v^7 \cr\cr
& & +27v^6+78v^4+q^3v^3+77v^5+110qv^3+56q^2v^2+89v^4q+33v^3q^2)
\label{f64}
\eeqs
\beq
f_{65}=-v^9(v+q)^4(v+1)^2(18v^2+11qv+7v^2q+3v^4+qv^3+14v^3+q^2v+2q^2)
\label{f65}
\eeq
\beq
f_{66}=v^{12}(v+q)^6(v+1)^4 \ .
\label{f66}
\eeq
The terms $\lambda_{Z,s3c,j}$ for $17 \le j \le 20$ are roots of the
quartic equation
\beq
\xi^4+f_{41}\xi^3+f_{42}\xi^2+f_{43}\xi+f_{44}=0
\label{eqquart}
\eeq
where
\beq
f_{41}=-(15v^3+6v^4+qv^3+12v^2q+5q^2v+q^3+v^5)
\label{f41}
\eeq
\beqs
& & f_{42}=v^2(v+q)(2q^3+32v^3+30v^2q+v^6+30v^4+8q^2v^2+13q^2v+4v^4q \cr\cr
& & +v^3q^2+q^3v+21qv^3+10v^5)
\label{f42}
\eeqs
\beqs
& & f_{43}=-v^4(v+1)(v+q)^2(q^3+15v^3+19v^2q+10v^4+4q^2v^2+9q^2v \cr\cr
& & +2v^4q+v^3q^2+10qv^3+2v^5)
\label{f43}
\eeqs
\beq
f_{44}=v^7(v+q)^5(v+1)^3 \ .
\label{f44}
\eeq

\subsection{Relations Between Potts Partition Function and Tutte Polynomial}

The formulas relating the Potts model partition function $Z(G,q,v)$ and the
Tutte polynomial $T(G,x,y)$ were given in \cite{a} and hence we shall be brief
here.  The Tutte polynomial of $G$, $T(G,x,y)$, is given by 
\cite{tutte1}-\cite{tutte3}
\beq
T(G,x,y)=\sum_{G^\prime \subseteq G} (x-1)^{k(G^\prime)-k(G)}
(y-1)^{c(G^\prime)}
\label{tuttepol}
\eeq
where $k(G^\prime)$, $e(G^\prime)$, and $n(G^\prime)=n(G)$ denote the number
of components, edges, and vertices of $G^\prime$, and
\beq
c(G^\prime) = e(G^\prime)+k(G^\prime)-n(G^\prime)
\label{ceq}
\eeq
is the number of independent circuits in $G^\prime$.
For the graphs of interest here, $k(G)=1$.  Now let
\beq
x=1+\frac{q}{v}
\label{xdef}
\eeq
and
\beq
y=a=v+1
\label{ydef}
\eeq
so that
\beq
q=(x-1)(y-1) \ .
\label{qxy}
\eeq
Then
\beq
Z(G,q,v)=(x-1)^{k(G)}(y-1)^{n(G)}T(G,x,y) \ .
\label{ztutte}
\eeq

For a planar graph $G$ the Tutte polynomial satisfies the duality relation
\beq
T(G,x,y) = T(G^*,y,x)
\label{tuttedual}
\eeq
where $G^*$ is the (planar) dual to $G$.  As discussed in \cite{a},
the Tutte polynomial for recursively defined graphs comprised of $m$
repetitions of some subgraph has the form
\beq
T(G_m,x,y) = \sum_{j=1}^{N_{T,G,\lambda}} c_{T,G,j}(\lambda_{T,G,j})^m \ .
\label{tgsum}
\eeq

\subsection{Cyclic and M\"obius $L_y=3$ Strips of the Square Lattice}

We calculate 
\beq
T(sq,3 \times m,FBC_y,PBC_x,x,y) = 
\sum_{j=1}^{20} c_{T,s3c,j}(\lambda_{T,s3c,j})^m
\label{tlxy}
\eeq
(do not confuse the $x$ and $y$ referring to the longitudinal and transverse
directions with the $x$ and $y$ variables defined in (\ref{xdef}) and
(\ref{ydef})) and
\beq
T(sq,3 \times m,FBC_y,TPBC_x,x,y) = 
\sum_{j=1}^{20} c_{T,s3Mb,j}(\lambda_{T,s3c,j})^m
\label{tmbxy}
\eeq
where
\beq
\lambda_{T,s3c,1} = 1
\label{lamtut1}
\eeq
\beq
\lambda_{T,s3c,2} = x
\label{lamtut2}
\eeq
\beq
\lambda_{T,s3c,(3,4)} = \frac{1}{2}\biggl [ x+y+3 \pm \sqrt{ (x+y+3)^2-4xy } 
\biggr ]
\label{lamtut34}
\eeq
and
\beq
\lambda_{T,s3c,(5,6)}= \frac{1}{2}\biggl [ x+y+1 \pm \sqrt{ (x+y+1)^2-4xy }
\biggr ] \ .
\label{lamtut56}
\eeq
Among the $\lambda_{T,s3c,j}$'s that depend on $x$ and $y$, four are symmetric
functions of these variables:
\beq
\lambda_{T,s3c,j}(x,y)=\lambda_{T,s3c,j}(y,x) \quad {\rm for} \ \ 3 \le j \le 6
 \ . 
\label{lams3tutsym}
\eeq
(Of course, $\lambda_{T,s3c,1}=1$ is trivially symmetric under $x
\leftrightarrow y$.) 
The $\lambda_{T,s3c,j}$ for $j=7,8,9$ are the roots of the cubic equation
\beq
\xi^3-(x+1)(x+y+1)\xi^2+xy(x^2+2x+y+2)\xi-y^2x^3=0 \ .
\label{eqcubtut}
\eeq
The $\lambda_{T,s3c,j}$ for $10 \le j \le 15$ are roots of the sixth degree
equation 
\beq 
\xi^6+f_{61T}\xi^5+f_{62T}\xi^4+f_{63T}\xi^3+f_{64T}\xi^2+f_{65T}\xi+f_{66T}=0
\label{eqsixtut}
\eeq
where
\beq
f_{61T}=-(xy+5+5x+2x^2+4y+y^2)
\label{f61tut}
\eeq
\beqs
& & f_{62T}=2x^2y^2+2x^3y+2+7y^2x+15yx+y^3x+10yx^2+12x^2+10x+x^4 \cr\cr
& & +6x^3+4y^2+y^3+5y
\label{f62tut}
\eeqs
\beqs
& & f_{63T}=-x(2y^3x^2+y^2x^3+20yx^2+9y+15y^2x+10x^2y^2+4y^3x+17yx \cr\cr
& & +x^4y+7x^3y+3+5x+8x^2+y^4+x^4+5x^3+11y^2+6y^3)
\label{f63tut}
\eeqs
\beqs
& & f_{64T}=yx^2(5y+8x+2+14yx+9x^2+y^2x^3+2y^3x+4x^2y^2+4y^2+y^3+x^4 \cr\cr
& & +5x^3+10yx^2+9y^2x+3x^3y)
\label{f64tut}
\eeqs
\beq
f_{65T}=-x^4y^2(3+4y+2y^2+y^2x+yx^2+x^2+3x+3yx)
\label{f65tut}
\eeq
\beq
f_{66T}=x^6y^4 \ .
\label{f66tut}
\eeq
The next term is 
\beq
\lambda_{T,s3c,16}=xy \ .
\label{lam16tut}
\eeq
Finally, the $\lambda_{T,s3c,j}$ for $17 \le j \le 20$ are roots of the quartic
equation 
\beq
\xi^4+f_{41T}\xi^3+f_{42T}\xi^2+f_{43T}\xi+f_{44T}=0
\label{eqquartut}
\eeq
where
\beq
f_{41T}=-(3+yx+4x+2x^2+3y+x^3+y^2)
\label{f41tut}
\eeq
\beq
f_{42T}=x(2y^2x+4x+3+4y^2+3yx^2+4yx+5y+x^3+3x^2+x^2y^2+x^3y+y^3)
\label{f42tut}
\eeq
\beq
f_{43T}=-yx^2(2x+1+2yx+2yx^2+2y+y^2+x^2y^2+x^3+3x^2)
\label{f43tut}
\eeq
\beq
f_{44T}=y^3x^5 \ .
\label{f44tut}
\eeq
The algebraic equations that yield the twenty terms can thus be summarized as
\beq
\{3(1), \ 2(2), \ 1(3), \ 1(4), \ 1(6) \} \quad {\rm for} \ \ sq, 3 \times L_x,
FBC_y,(T)PBC_x
\label{cycmbstructure}
\eeq
by which we mean that these equations consist of three linear, two quadratic,
one cubic, one quartic, and one sixth-degree equation.  This is compared with
the results for the Potts model partition functions for the corresponding 
strips of the square lattice with $L_y=1$ and 2 in Table \ref{proptable}.

\begin{table}
\caption{\footnotesize{Some properties of $Z(G,q,v)$ for strips of the square
lattice of width $L_y$ and arbitrary length, with periodic longitudinal 
boundary conditions.}}
\begin{center}
\begin{tabular}{|c|c|c|c|c|c|c|}
\hline\hline 
$Z$ or $P$&$L_y$ &$BC_y$& $BC_x$ & $N_\lambda$ & eqs.   & ref. \\ \hline\hline
$Z(G,q,v)$ & 1 & $FBC_y$ & $(T)PBC_x$ & 2  & \{2(1)\}    & $-$   \\ \hline
           & 2 & $FBC_y$ & $(T)PBC_x$ & 6  & \{2(1),2(2)\} & [14]  \\ \hline
           & 3 & $FBC_y$ & $(T)PBC_x$ & 20 & \{3(1),2(2),1(3),1(4),1(6)\} &
here \\ \hline
           & 2 & $PBC_y$ & $(T)PBC_x$ & 6  & \{2(1),2(2)\} & here \\ \hline 
           & 3 & $PBC_y$ & $PBC_x$    & 20 & \{4(1),3(2),2(3),1(4)\} & here 
\\ \hline
           & 3 & $PBC_y$ & $TPBC_x$   & 12 & \{3(1),1(2),1(3),1(4)\} & here \\ 
\hline\hline
$P(G,q)$   & 1 & $FBC_y$ & $(T)PBC_x$ & 2  & \{2(1)\} & $-$        \\ \hline
           & 2 & $FBC_y$ & $(T)PBC_x$ & 4  & \{4(1)\} & [25,24]    \\ \hline
           & 3 & $FBC_y$ & $(T)PBC_x$ & 10 & \{5(1),1(2),1(3)\} & [46,47,50]
 \\ \hline
           & 4 & $FBC_y$ & $(T)PBC_x$ & 26 & \{4(1),1(2),2(3),1(4),2(5)\} 
& [54] \\ \hline
           & 3 & $PBC_y$ & $PBC_x$    & 8 & \{8(1)\} & [51] \\ \hline
           & 3 & $PBC_y$ & $TPBC_x$   & 5 & \{5(1)\} & [51] \\ \hline
           & 4 & $PBC_y$ & $PBC_x$    & 33& \{9(1),6(2),4(3)\} & [56] \\ \hline
           & 4 & $PBC_y$ & $TPBC_x$   & 22& \{7(1),3(2),3(3)\} & [56] \\ 
\hline\hline
\end{tabular}
\end{center}
\label{proptable}
\end{table}

It is convenient to extract a common factor from the coefficients:
\beq
c_{T,G,j} \equiv \frac{\bar c_{T,G,j}}{x-1} \ . 
\label{cbar}
\eeq
Of course, although the individual terms contributing
to the Tutte polynomial are thus rational functions of $x$ rather than
polynomials in $x$, the full Tutte polynomial is a polynomial
in both $x$ and $y$.  We have
\beq
\bar c_{T,G,j}=c_{Z,G,j} \quad \forall \ j \ .
\label{cj}
\eeq

  Our calculations are in accord with the following generalization for the
family of strips denoted by $G=sq, L_y \times L_x, FBC_y, (T)PBC_x$:
\beq
\lambda_{T,G,1}=1
\label{lam1tutgeneral}
\eeq
\beq
\lambda_{Z,G,2}=x
\label{lam2tutgeneral}
\eeq
with coefficients for the cyclic strips
\beq
\bar c_{T,G_{cyc.},1}=c_{Z,G_{cyc.},1}=c^{(L_y)}
\label{c1tutgeneral}
\eeq
\beq
\bar c_{T,G_{cyc.},2}=c_{Z,G_{cyc.},2}=c^{(L_y-1)} \ . 
\label{c2tutgeneral}
\eeq
For the M\"obius strips, the coefficients are determined from the general
formulas given in \cite{cf}. 

\subsection{$L_y=2$ Strips of the Square Lattice with Torus and Klein Bottle
Boundary Conditions}

We calculate the Tutte polynomials
\beq
T(sq,2 \times m,PBC_y,PBC_x,x,y) =
\sum_{j=1}^6 c_{T,s2t,j}(\lambda_{T,s2t,j}(x,y))^m
\label{tgsums2torus}
\eeq
and
\beq
T(sq,2 \times m,PBC_y,TPBC_x,x,y) =
\sum_{j=1}^6 c_{T,s2k,j}(\lambda_{T,s2t,j}(x,y))^m
\label{tgsums2klein}
\eeq
where 
\beq
\lambda_{T,s2t,1}=1
\label{lamtor2tut_1}
\eeq
\beq
\lambda_{T,s2t,2}=x
\label{lamtor2tut_2}
\eeq
\beq
\lambda_{T,s2t,(3,4)}=\frac{1}{2}\biggl [ x+2+2y+y^2 \pm \sqrt{R_{T234}} \ 
\biggr ]
\label{lamtor2tut_34}
\eeq
where
\beq
R_{T234}= (x+2+2y+y^2)^2-4xy^2
\label{tr223}
\eeq
\beq
\lambda_{T,s2t,(5,6)}=\frac{1}{2}\biggl [x^2+xy+x+1+y+y^2 \pm 
\sqrt{R_{T256}} \ \biggr ] 
\label{lamtor2tut_56}
\eeq
where
\beq
R_{T256}=(y^2+y+1+3xy+x+x^2)(y^2+y+1-xy+x+x^2) \ . 
\label{rt256}
\eeq
Note that
\beq
\lambda_{T,s2t,j}(x,y) = \lambda_{T,s2t,j}(y,x) \quad {\rm for} \ \ j=5,6
\ .
\label{lam5tor2sym}
\eeq
The corresponding reduced coefficients (defined as in eq. (\ref{cbar})) are
given by (\ref{c1tor2})-(\ref{c56kb2}).

\subsection{$L_y=3$ Strips of the Square Lattice with Torus and Klein Bottle
Boundary Conditions} 

We calculate the Tutte polynomials 
\beq
T(sq,3 \times m,PBC_y,PBC_x,x,y) =
\sum_{j=1}^{20} c_{T,s3t,j}(\lambda_{T,s3t,j}(x,y))^m
\label{tgsums3torus}
\eeq
and
\beq
T(sq,3 \times m,PBC_y,TPBC_x,x,y) =
\sum_{j=1}^{12} c_{T,s3k,j}(\lambda_{T,s3k,j}(x,y))^m
\label{tgsums3klein}
\eeq
where (ordering the $\lambda_{T,s3t,j}$'s by decreasing degree of the
associated coefficients $c_{T,s3t,j}$)
\beq
\lambda_{T,s3t,j}=\lambda_{T,s3,j} \quad {\rm for} \ \ 1 \le j \le 6
\label{lamtor1to6tut}
\eeq
\beq
\lambda_{Z,s3t,(7,8)}=\frac{1}{2}\biggl [ x+y+4 \pm \sqrt{(x+y+4)^2-4xy}
\biggr ]
\label{lamtor78tut}
\eeq
\beq
\lambda_{T,s3t,9}=y
\label{lamtor9tut}
\eeq
\beq
\lambda_{T,s3t,10}=xy
\label{lamtor10tut}
\eeq
and
\beq
\lambda_{T,s3t,j}=\lambda_{T,s3,j-4} \quad {\rm for} \ \ j=11,12,13 \ . 
\label{lamtor14to16tut}
\eeq
The $\lambda_{T,s3t,j}$ for $14 \le j \le 17$ are roots of the quartic
\beq
\xi^4+g_{41T}\xi^3+g_{42T}\xi^2+g_{43T}\xi+g_{44T}=0
\label{eqquartortut}
\eeq
where
\beq
g_{41T}=-(7+yx+y^3+x^2+5x+7y+3y^2)
\label{g41t}
\eeq
\beq
g_{42T}=y(8x+6y+6+4y^2+x^3+y^3+y^3x+3yx^2+5x^2+7yx+5y^2x+y^2x^2)
\label{g42t}
\eeq
\beq
g_{43T}=-y^2x(3x+x^2+3yx+3+6y+5y^2+y^3+2y^2x+y^2x^2)
\label{g43t}
\eeq
\beq
g_{44T}=y^5x^3 \ .
\label{g44t}
\eeq
Finally, the $\lambda_{T,s3t,j}$ for $j=18,19,20$ are roots of the cubic 
equation
\beq
\xi^3+g_{31T}\xi^2+g_{32T}\xi+g_{33T}=0
\label{eqcubtortut}
\eeq
where
\beq
g_{31T}=-\Bigl [4+5(x+y)+3(x^2+y^2)+2xy+x^3+y^3 \Bigr ]
\label{g31t}
\eeq
\beq
g_{32T}=xy\Bigl [1+2(x+y)+3(x^2+xy(x+y)+y^2)+4xy+(x^3+y^3)+x^2y^2 \Bigr ]
\label{g32t}
\eeq
\beq
g_{33T}=-y^4x^4 \ .
\label{g33t}
\eeq
Among the $\lambda_{T,s3t,j}$'s that depend on $x$ and $y$, ten are symmetric
functions of these variables:
\beq
\lambda_{T,s3t,j}(x,y)=\lambda_{T,s3t,j}(y,x) \quad {\rm for} \ \
j=3,4,5,6,7,8,10,18,19,20 \ . 
\label{lamtorsym}
\eeq
(and $\lambda_{T,s3t,1}=1$ is trivially symmetric under $x
\leftrightarrow y$.)
The terms that enter in eq. (\ref{tgsums3klein}) for the strip with Klein
bottle boundary conditions are 
\beq
\lambda_{T,s3k,j}=\lambda_{T,s3t,j} \quad {\rm for} \ \ j=1,2
\label{lamklein12tut}
\eeq
\beq
\lambda_{T,s3k,j}=\lambda_{T,s3t,j+4}  \quad {\rm for} \ \ 3 \le j \le 5
\label{lamklein3to5tut}
\eeq
\beq
\lambda_{T,s3k,j}=\lambda_{T,s3t,j+8} \quad {\rm for} \ \ 6 \le j \le 12 \
.
\label{lamklein6to12tut}
\eeq
The eight terms in the Tutte polynomial for the torus strip 
$\lambda_{T,s3t,j}$ with $3 \le j \le 6$ and $10 \le j \le 13$ do not
occur in the Tutte polynomial for the Klein bottle strip.

For both of the types of strips $G_s=sq,3 \times m, PBC_y,(T)PBC_x$, 
the corresponding coefficients satisfy
\beq
{\bar c}_{T,G_s,j}=c_{Z,G_s,j} \quad \forall \ j \ .
\label{ctztor}
\eeq

\subsection{Special Values of Tutte Polynomials for $L_y=3$ Strips of the 
Square Lattice}

\subsubsection{General Relations}

For a given graph $G=(V,E)$, at certain special values of the arguments $x$ and
$y$, the Tutte polynomial $T(G,x,y)$ yields quantities of basic graph-theoretic
interest.  We recall some definitions: a spanning
subgraph was defined at the beginning of the paper; a tree is a
connected graph with no cycles; a forest is a graph containing one or
more trees; and a spanning tree is a spanning subgraph that is a tree.  We
recall that the graphs $G$ that we consider are connected.  Then the number
of spanning trees of $G$, $N_{ST}(G)$, is
\beq
N_{ST}(G)=T(G,1,1) \ ,
\label{t11}
\eeq
the number of spanning forests of $G$, $N_{SF}(G)$, is
\beq
N_{SF}(G)=T(G,2,1) \ ,
\label{t21}
\eeq
the number of connected spanning subgraphs of $G$, $N_{CSSG}(G)$, is
\beq
N_{CSSG}(G)=T(G,1,2) \ ,
\label{T12}
\eeq
and the number of spanning subgraphs of $G$, $N_{SSG}(G)$, is
\beq
N_{SSG}(G)=T(G,2,2) \ .
\label{t22}
\eeq
An elementary theorem (e.g., \cite{ka}) is that 
\beq
N_{SSG}(G)=2^{e(G)} \ . 
\label{nssggen}
\eeq

Since $T(G,L_y \times m,x,y)$ grows exponentially as $m
\to \infty$ for lattice strips $G$ 
of the type considered here, for the values 
$(x,y)=(1,1)$, (2,1), (1,2), and (2,2), one defines the corresponding 
constants
\beq
z_{set}(\{G\}) = \lim_{n(G) \to \infty} n(G)^{-1} \ln N_{set}(G) \ , \quad
set = ST, \ SF, \ CSSG, \ SSG
\label{zset}
\eeq
where, as above, the symbol $\{G\}$ denotes the limit of the graph family $G$
as $n(G) \to \infty$. These constants are determined completely by the
term $\lambda_{Z,G,j}$ that is dominant in the PM phase.  Since this is the
same for a given $G$ with a specified transverse boundary conditions,
independent of the longitudinal boundary conditions, the resultant value of
$z_{set}$ is the same for a strip of a given lattice type $\Lambda$, width
$L_y$, and transverse boundary condition, independent of the longitudinal
boundary condition ($FBC_x, PBC_x, TPBC_x$) \cite{a}. 

\subsubsection{Free Transverse Boundary Conditions}

For the $L_y=3$ cyclic and M\"obius strips of the square lattice we thus have
\beq
z_{ST,3 \times \infty,FBC_y}
=\frac{1}{3}\ln [\lambda_{T,s3c,17,x=1,y=1}] = 0.84307429...
\label{zsts3}
\eeq
As a special case of the result noted above, this is the same for the open
$L_y=3$ strip of the square lattice. 

A general upper bound on the number of spanning trees for a graph $G$ is 
\cite{grim}
\beq
N_{ST}(G) \le \frac{1}{n}\biggl ( \frac{2|E|}{n-1}\biggr )^{n-1} \ . 
\label{kboundg}
\eeq
For $n \to \infty$, in terms of the effective degree $\Delta_{eff}$ defined in
(\ref{delta_eff_cyc}), this yields 
\beq
z_{ST,\{G\}} \leq \ln \Delta_{eff} \ .
\label{b1nonreg}
\eeq
That is, for the cyclic, M\"obius, or free strip of the square lattice with 
width $L_y$, 
\beq
z_{ST,sq,L_y \times \infty,FBC_y,} \leq \ln \biggl [4 \Bigl (1-\frac{1}{2L_y} 
\Bigr ) \biggr ] \ .
\label{zstcycupper}
\eeq
For $L_y=3$, this evaluates to
\beq
z_{ST,sq,3 \times \infty,FBC_y} \leq \ln(10/3) = 1.2039728...
\label{zstcycupperly3}
\eeq
The ratio of the actual $z_{ST}$ to this upper bound is thus
\beq
\frac{z_{ST,sq,3 \times \infty,FBC_y}}{\ln(10/3)} = 0.7002436...
\label{rstcyc}
\eeq
We may also compare the value $z_{ST}$ in (\ref{zsts3}) with the value for the 
$L_y=2$ cyclic, M\"obius, or free strips of the square lattice, namely (e.g.,
\cite{a}) 
\beq
z_{ST,sq,2 \times \infty,FBC_y}= \frac{1}{2}\ln(2+\sqrt{3}) 
= 0.6584789...
\label{zstlad}
\eeq
Thus, as one increases the width of the strip with $FBC_y$ from $L_y=2$ to
$L_y=3$, the value of $z_{ST}$ increases about 30 \%. 
Another comparison of interest is the ratio of $z_{ST}$ for these finite-width
strips with $z_{ST}$ for the full 2D square lattice, which has the value
\cite{temprel,wu77}
\beq
z_{ST,sq} = \frac{4}{\pi}G_{Cat.} = 1.1662436...
\label{zsq}
\eeq
where $G_{Cat.}$ is the Catalan constant, 
\beq
G_{Cat.}=\sum_{j=0}^\infty \frac{(-1)^j}{(2j+1)^2} = 0.91596559...
\label{catalan}
\eeq
We have 
\beq
\frac{z_{ST,sq,2 \times \infty,FBC_y}}{z_{ST,sq}}=0.564615...
\label{rladdersq}
\eeq
and
\beq
\frac{z_{ST,sq,3 \times \infty,FBC_y}}{z_{ST,sq}}=0.722897...
\label{rs3sq}
\eeq
Thus, for $L_y=3$, the value of $z_{ST,sq,L_y \times
\infty,FBC_y}$ is within about 30 \% of the value (\ref{zsq}) for the 
infinite square lattice. 

Our calculations also yield, for spanning forests, 
\beq
z_{SF,3 \times \infty,FBC_y}=\frac{1}{3}\ln [\lambda_{T,s3c,17,x=2,y=1}] 
=1.106676...
\label{zsfs3}
\eeq
and, for the number of connected spanning subgraphs, 
\beq
z_{CSSG,3\times \infty,FBC_y}=\frac{1}{3}\ln [\lambda_{T,s3c,17,x=1,y=2}]=
0.966005...
\label{zcssgs3}
\eeq

 From the general theorem (\ref{nssggen}), together with the definition 
(\ref{deltaeff}), we have 
\beq
z_{SSG}=\frac{\Delta_{eff}}{2}\ln 2 \ .
\label{zssggen}
\eeq
In the present case, with $\Delta_{eff}=10/3$ for the $L_y=3$ cyclic or
M\"obius strip graph, we have 
\beq
z_{SSG}=\frac{5}{3}\ln 2 =1.155245...
\label{zssgcyc}
\eeq

\subsubsection{Periodic Transverse Boundary Conditions}

For the $L_y=2$ and $L_y=3$ strips of the square lattice with periodic 
transverse boundary conditions (torus, Klein bottle and cylindrical b.c.) we 
have
\beq
z_{ST,2 \times \infty,PBC_y} =
\frac{1}{2}\ln [\lambda_{T,s2t,5,x=1,y=1}] 
= \frac{1}{2}\ln (3 + 2\sqrt{2} \ ) = 0.8813735...
\label{zst_tor2}
\eeq
\beqs
z_{ST,3 \times \infty,PBC_y} & = & 
\frac{1}{3}\ln [\lambda_{T,s3t,18,x=1,y=1}] 
= \frac{1}{3}\ln \biggl [ \frac{1}{2}\Bigl (23+5\sqrt{21} \Bigr ) \biggr ] 
\cr\cr & = & 
  \frac{2}{3}\ln \biggl [ \frac{1}{2}\Bigl (5 + \sqrt{21} \Bigr ) \biggr ]
= 1.0445328...
\label{zsttor3}
\eeqs
These results agree with the eq. (6.6.1) in \cite{st}, where a general
discussion was given for the number of spanning trees on various lattices and
lattice strips. 
The torus is a $\Delta$-regular graph and, even for the cylindrical strip, in
the $L_x \to \infty$ limit, the effect of the end vertices goes to zero, so
that in all cases, the coordination number is 4.  Hence, the upper bound
(\ref{kboundg}) reads 
\beq
z_{ST,sq,L_y \times \infty,PBC_y} \leq \ln 4 \ .
\label{zstsqcyl}
\eeq
The ratios of the actual values of $z_{ST}$ for $L_y=2$ and $L_y=3$ to this 
upper bound are thus
\beq
\frac{z_{ST,sq,2 \times \infty,PBC_y}}{\ln 4} = 0.635777...
\label{rstcyl2}
\eeq
\beq
\frac{z_{ST,sq,3 \times \infty,PBC_y}}{\ln 4} = 0.753471...
\label{rstcyl3}
\eeq
Although the $L_y=3$ strips with free transverse boundary conditions are not
$\Delta$-regular graphs, the corresponding strips with periodic transverse
boundary conditions (and, say, periodic longitudinal boundary conditions) are
$\Delta$-regular, so 
one may also compare the exact results with a sharper upper
bound for a $\Delta$-regular graph $G$ with $n$ vertices {\cite{mckay,cy}, 
 \beq
N_{ST}(G) \le \Biggl ( \frac{2\ln n}{n \Delta \ln \Delta}
\Bigg) (C_\Delta) ^n
\label{nmckay}
\eeq
where
\beq
C_\Delta = \frac{(\Delta-1)^{\Delta-1}}{[\Delta(\Delta-2)]^{\Delta/2-1}} 
\label{ck}
\eeq
which yields 
\beq
z_{\{G\}} \leq z_{\{G\},\Delta} = \ln C_\Delta
\label{zmckay} 
\eeq
where we label the upper bound for the limit of $\Delta$-regular graphs as 
$z_{\{G\},\Delta}$. We have
\beq
\frac{z_{ST,2 \times \infty,PBC_y}}{3\ln(3/2)}=0.724578...
\label{ratioregular2}
\eeq
\beq
\frac{z_{ST,3 \times \infty,PBC_y}}{3\ln(3/2)}=.8587116... 
\label{ratioregular3}
\eeq
Comparing with the value of $z_{ST}$ for the 2D square lattice (cf. eq. 
(\ref{zsq}) above), we have 
\beq
\frac{z_{ST,sq,2 \times \infty,PBC_y}}{z_{ST,sq}}=0.755737...
\label{rtor2st}
\eeq
\beq
\frac{z_{ST,sq,3 \times \infty,PBC_y}}{z_{ST,sq}}= 0.8956386...
\label{rtor3st}
\eeq
It is impressive that the strips with rather modest $L_y=3$ width and
periodic transverse boundary conditions (and any longitudinal boundary
conditions) yield a value of $z_{ST,sq,L_y \times \infty,PBC_y}$ that is only 
about 10 \% below the value for the infinite 2D square lattice.  

We also calculate
\beq
z_{SF,2 \times \infty,PBC_y}=\frac{1}{2}\ln [\lambda_{T,s2t,5,x=2,y=1}] 
= \frac{1}{2}\ln \biggl [ \frac{1}{2}(11+\sqrt{105} \ )\biggr ] = 1.181533...
\label{zsfst2}
\eeq
\beq
z_{SF,3 \times \infty,PBC_y}=\frac{1}{3}\ln [\lambda_{T,s3t,18,x=2,y=1}]
= 1.263294...
\label{zsfst3}
\eeq
\beq
z_{CSSG,2\times \infty,PBC_y}=\frac{1}{2}\ln [\lambda_{T,s2t,5,x=1,y=2}]
= \frac{1}{2}\ln \biggl [ \frac{1}{2}(11+\sqrt{105} \ )\biggr ] = 1.181533...
\label{zcssgst2}
\eeq
and 
\beq
z_{CSSG,3\times \infty,PBC_y}=\frac{1}{3}\ln [\lambda_{T,s3t,18,x=1,y=2}]=
1.263294...
\label{zcssgst3}
\eeq
The equality of (\ref{zcssgst2}) with (\ref{zsfst2}) and the equality 
of (\ref{zcssgst3}) with (\ref{zsfst3}) follow from the $x \leftrightarrow y$
symmetry of $\lambda_{T,s2t,5}$ and $\lambda_{T,s3t,18}$, as indicated above in
eqs. (\ref{lam5tor2sym}) and (\ref{lamtorsym}). Finally, from 
eq. (\ref{zssggen}) with $\Delta = 4$, we have, for both the torus or Klein
bottle strip with arbitrary $L_y$, 
\beq
z_{SSG}=2 \ln 2 =1.386294...
\label{zssgtorus}
\eeq

\vfill
\eject
\end{document}